
\input harvmac
\input epsf.tex
\def\caption#1{{\it
	\centerline{\vbox{\baselineskip=12pt
	\vskip.15in\hsize=4.5in\noindent{#1}\vskip.1in }}}}
\def\pyidk{PHY-9057135}
\def\dint{\int \kern-.6em \int\kern-.2em}
\def\iso{${ }^1S_0$}
\def\({\left(}
\def\){\right)}
\def\CJ{{\cal J}}
\def\CK{{\cal K}}
\def\CM{{\cal M}}
\def\MeV{{\rm MeV}}
\def\bfq{{\bf q}}

\def\bfp{{\bf p}}
\def\bfpp{{\bf p'}}
\def\kcd{|\bfp|\cot\delta(\bfp)}
\def\too#1{\,\mathop{\longrightarrow}\limits_{#1}\, }
\def\bar#1{\overline{#1}}
\def\ms{$\bar {MS}$}

\def\bra#1{\left\langle #1\right|}
\def\ket#1{\left| #1\right\rangle}

\def\half{{\textstyle{1\over2}}} 
\def\frac#1#2{{\textstyle{#1\over #2}}}

\def\igralt#1{\int{{\rm d}^3{\bf #1}\over (2\pi)^3}\,}
\def\igraln#1{\int{{\rm d}^n{\bf #1}\over (2\pi)^n}\,}
\def\digraln{\dint{{\rm d}^n{\bfq}\over (2\pi)^n}{{\rm d}^n{\bfq'}\over
(2\pi)^n}\,}
\def\igralf#1{\int{{\rm d}^4{ #1}\over (2\pi)^4}\,}
\def\msb{{\mathop{{\bar {MS}}}}}
%
%
\def\ltap{\ \raise.3ex\hbox{$<$\kern-.75em\lower1ex\hbox{$\sim$}}\ }
\def\gtap{\ \raise.3ex\hbox{$>$\kern-.75em\lower1ex\hbox{$\sim$}}\ }
\def\gl{\ \raise.5ex\hbox{$>$}\kern-.8em\lower.5ex\hbox{$<$}\ }
\def\roughly#1{\raise.3ex\hbox{$#1$\kern-.75em\lower1ex\hbox{$\sim$}}}
\def\ie{\hbox{\it i.e.}}        
\def\eg{\hbox{\it e.g.}}

\def\np#1#2#3{{Nucl. Phys. } B{#1} (#2) #3}
\def\pl#1#2#3{{Phys. Lett. } {#1}B (#2) #3}
\def\prl#1#2#3{{Phys. Rev. Lett. } {#1} (#2) #3}
\def\physrev#1#2#3{{Phys. Rev. } {#1} (#2) #3}

\relax
\def\Dsl{\,\raise.15ex \hbox{/}\mkern-13.5mu D}
\def\[{\left[}
\def\]{\right]}
\def\({\left(}
\def\){\right)}
\noblackbox
\def\pyidk{PHY-9057135}
\font\ninerm=cmr9
\def\Title#1#2{\nopagenumbers\abstractfont\hsize=\hstitle\rightline{{\ninerm
#1}}
\vskip .4in\centerline{\titlefont #2}\abstractfont\vskip .5in\pageno=0}
\Title{\vbox{
\hbox{DOE/ER/40561-257-INT96-00-125 }
\hbox{UW/PT 96-06}
\hbox{CMU-HEP96-06 }
\hbox{DOE-ER-40862-117}
\hbox{CALT-68-2047} }}
{\vbox{\centerline{Nucleon-Nucleon Scattering}
\bigskip
\centerline{from  Effective Field Theory }}}
\vskip-.2in
\centerline{David B. Kaplan}
\centerline{{\sl
Institute for Nuclear Theory, University of Washington}}
\centerline{{\sl Box 351550, Seattle WA 98195-1550}}
\centerline{{\tt dbkaplan@phys.washington.edu}}
\medskip
\medskip
\centerline{  Martin J. Savage
\footnote{$^{\dagger}$}{  DOE Outstanding Junior Investigator.  Address after
Sept. 1,  1996: Dept. of Physics, University of Washington, Box 351550, Seattle
WA 98195-1550.}}
\centerline{{\sl   Department of Physics, Carnegie Mellon University,
Pittsburgh PA 15213}}
\centerline{
{\tt   savage@thepub.phys.cmu.edu}}
\medskip
\medskip
\centerline{  Mark B. Wise}
\centerline{{\sl   California Institute of Technology, Pasadena, CA
91125}}
\centerline{{\tt   wise@theory.caltech.edu}}
\bigskip\bigskip\vfill
{ We perform a nonperturbative calculation of the \iso\ $NN$ scattering
amplitude, using an effective field theory (EFT) expansion. The expansion we
advocate is a modification of what has been used previously; it is not a chiral
expansion in powers of $m_\pi$.  We use dimensional
regularization throughout, and the $\msb$ renormalization scheme; our final
result depends only on physical observables.  We show
that the EFT expansion of the quantity  $\kcd$ converges at momenta much
greater than the scale $\Lambda$ that characterizes the derivative expansion of
the EFT Lagrangian. Our conclusions are optimistic about the applicability of
an EFT approach to the quantitative study of nuclear matter.}
\Date{5/96}
\baselineskip 18pt

\newsec{Introduction}

Effective field theories are routinely used in particle physics and have proved
an invaluable tool for computing physical quantities in theories with disparate
energy scales
\ref\eft{
S. Weinberg, Physica (Amsterdam) 96A  (1979) 327;
E. Witten, \np{122}{1977}{109}.}
(for recent reviews see
\ref\eftrev{
H. Georgi, Ann. Rev. of Nucl. and Part. Sci., 43 (1993) 209;
J. Polchinski, Proceedings of  {\it Recent Directions in Particle Theory} ,
TASI92
(1992) 235;
A. V. Manohar, {\it Effective Field Theories}, hep-ph/9508245;
D.B. Kaplan,  {\it Effective Field Theories}, hep-ph/9506035.
}).
Several years ago,
Weinberg proposed that the machinery of effective field theory (EFT) could be
applied fruitfully to nucleon-nucleon scattering and nuclear physics
\ref\weinberg{
 S. Weinberg, \pl{251}{1990}{288};
\np{363}{1991}{3}; \pl{295}{1992}{114}.}.
Nucleon interactions might be
profitably treated by EFT since they involve several different physical scales,
such as the nucleon mass $M$, the pion and vector meson masses ($m_\pi$,
$m_\rho$, $m_\omega$, etc.).
Furthermore,
chiral symmetry in nucleon-pion interactions is necessarily expressed in the
language of EFT, and the chiral expansion around $m_\pi=0$ gives one a natural
expansion parameter.  Since Weinberg's original papers, much work has been done
in the subject, with fair success in reproducing low energy features of
nucleon-nucleon scattering from a chiral Lagrangian description of nucleon
interactions \ref\kolcka{
C. Ordonez, U. van Kolck, \pl{291}{1992}{459};
C. Ordonez, L. Ray, U. van Kolck, \prl{72}{1994}{1982};
Phys. Rev. C 53 (1996) 2086, nucl-th/9511380.}
\ref\kolckb{ U. van Kolck, \physrev{C49}{1994}{2932}.}.

The goal of an EFT description of nuclear physics is not to improve upon
semi-phenomenological models of the nucleon-nucleon interaction, such as the
Paris
\ref\paris{W. N. Cottingham {\it et al}, \physrev{D8}{1973}{800}.} ,
Bonn
\ref\bonn{R. Machleidt, K. Holinde and C. Elster, Phys. Rep. 149 (1987) 1.}
or Nijmegen
\ref\nijmegen{M.M. Nagels, T.A. Rijken and J.J. de Swart,
\physrev{D17}{1978}{768}.}
potentials .
Instead it has been used to relate 3-body forces to 2-body forces
\weinberg\kolckb and   to
explain the observed hierarchy of isospin violation
\ref\kolckc{ U. van Kolck, J.L. Friar and  T. Goldman, \pl{371}{1996}{169}.}.
One can also
investigate the role of strangeness in hypernuclei or dense matter, along the
lines of
\ref\SaWia{M.J. Savage and M.B. Wise, \physrev{D53}{1996}{349}.}.
More generally, it allows one to better
understand the physical origin of various features of the nucleon interaction
(for recent progress in this direction see
\nref\pmr{
T.-S. Park, D.-P. Min and M. Rho, Phys. Rept. 233 (1993) 341;
T.-S. Park, I. S. Towner and K. Kubodera, Nucl. Phys. A579
(1994) 381;
T.-S. Park, D.-P. Min and M. Rho, Phys.Rev.Lett.74 (1995) 4153;
T.-S. Park, D.-P. Min and M. Rho, Nucl.Phys.A596 (1996) 515.}
\nref\friara{J.L. Friar, D.G. Madland and B.W Lynn,
nucl-th/9512011 (1995).}\nref\friarb{J.L. Friar,
{\it Nuclear Forces and Chiral Theories},
Few-Body Systems Suppl. 99 (1996) 1.}\nref\kolckrev{
U. van Kolck, {\it Effective Chiral Theory of Nuclear Forces},
Lectures presented at the 7th Summer School and Symposium
on Nuclear Physics, Seoul,
(1994).}\refs{\pmr - \kolckrev}\
and references therein).
One may also hope that the technique will allow semi-analytical
approaches to solving 2- and many-body problems now only approached
numerically.

A fundamental difficulty in an EFT description of nuclear forces is
that they are necessarily nonperturbative, so that an infinite series of
Feynman diagrams must be summed.  Which diagrams must be summed is well known,
and the summing them is equivalent to solving a Schr\"odinger equation.
However,
an EFT yields graphs which require renormalization, giving rise to a
Schr\"odinger potential which is too singular to solve conventionally.  In
Weinberg's work  \weinberg, only a contact interaction was summed, and the
system was renormalized; in
\kolcka\kolckb\kolckc,
more complicated interactions are
considered and a momentum cut-off is implemented, with bare couplings chosen to
best fit phase shift data.  In this paper we focus on the \iso\ ($np$) partial
wave, and show how to compute the phase shift beyond lowest order in the EFT
expansion, using dimensional regularization and the minimal subtraction
($\msb$) renormalization scheme.

Another problem with discussing systems with barely bound
(or nearly bound)  states in the language of EFT
is that a new length scale emerges that is not directly associated with any
physical threshold --- the scattering length $a$.  This makes
the power counting in an EFT with large scattering length much less obvious
than in one without.    \iso\ nucleon-nucleon scattering is particularly
problematic from an effective field theory point of view, since the scattering
length is very  large: $a\sim -24\ {\rm fm}\ \sim (8.5\
\MeV)^{-1}$, a mass scale far lower than any hadron mass. In
this paper we propose a specific ordering of the EFT expansion
to avoid this problem.  
In the process, we are led to a modification of the
power counting scheme proposed by Weinberg.

We begin by briefly reviewing Weinberg's power counting scheme and the
connection between Feynman diagrams and the Schr\"odinger equation.  We then
show
how to sum the relevant graphs even when they are divergent, and we construct
the low energy EFT for nucleons alone in the \ms\ scheme.  Finally we construct
the EFT including one pion exchange in the \ms\ scheme; in both cases we show
at what scale the EFT fails.
We conclude with thoughts about improving the
approach, and its applicability to finite density calculations.

\newsec{Effective field theory, power counting and the Schr\"odinger equation}

\subsec{Weinberg's power counting scheme}
The philosophy of EFT is that for scattering processes
involving external momenta $\ltap Q$, one need only consider a Lagrangian which
explicitly includes light degrees of freedom for which $m\ltap Q$.  The effects
of heavy virtual particles appear as an infinite number of
nonrenormalizable operators suppressed by powers of the mass scale $\Lambda$
relevant to the degrees of freedom excluded from the theory.  EFT's can be
predictive since amplitudes may be expanded in powers of $Q/\Lambda$, so that
the effect of a nonrenormalizable operator on low energy physics is less
important the higher the dimension of that operator.

The scale $\Lambda$ can be determined by fitting low energy data to the
predictions of the EFT to sufficient accuracy;  the lower the scale $\Lambda$,
the smaller the momentum range over which the EFT is predictive.
An EFT will have to be modified as one approaches $Q\simeq \Lambda$, and the
degrees of freedom with mass $\Lambda$ must then be explicitly included in the
EFT.   One then has a new EFT characterized by a scale $\Lambda'$, which
characterizes the next level of particles excluded from the theory.  An EFT is
only useful to the extent that there is a well defined hierarchy of mass
scales; if there is such a hierarchy one can typically predict a
large amount of data in terms of a few parameters.

A necessary ingredient for an EFT is a power counting scheme that tells one
what graphs to compute to any order in the momentum expansion.  We  reproduce
here Weinberg's analysis for $NN$ scattering, couched however in the language
of covariant rather than time ordered  perturbation theory.  The main
complication arises from the fact that a nucleon propagator $S(q)=i/(q_0 - {\bf
q}^2/2M)$ scales like $1/Q$ if $q_0$ scales like $m_\pi$ or an external
3-momentum,    while $S(q)\sim M/Q^2$ if $q_0$ scales like an external kinetic
energy.  Similarly, in loops $\int {\rm d} q_0$ can scale like $Q$ or $Q^2/M$,
depending on which type of pole is picked up. To distinguish between these two
scaling properties
we begin  by defining generalized ``$n$-nucleon potentials'' $V^{(n)}$
comprised of those parts of connected Feynman diagrams with
$2n$ external nucleon lines
that have no powers of $M$ in their scaling \foot{With the exception of inverse
powers of $M$ from relativistic corrections.}. Such a  diagram always has
exactly $n$ nucleon lines running through it, since there is no
nucleon-antinucleon pair creation in the effective theory.  $V^{(n)}$ includes
(i) diagrams which are $n$-nucleon irreducible; (ii) parts of diagrams which
are 1-nucleon irreducible\foot{An $n$-nucleon irreducible diagram is one which
does not fall apart when $n$ nucleon lines are cut.}.
To compute the latter contribution  to $V^{(n)}$
one identifies all combinations of two or more internal nucleon lines
that can be simultaneously on-shell, and excludes their pole
contributions when performing the $\int{\rm d}q_0$ loop integrations.  An
example of the 2-pion exchange contributions to $V^{(2)}$ is shown in Fig.~1.
\topinsert
\centerline{\epsfxsize=4in\epsfbox{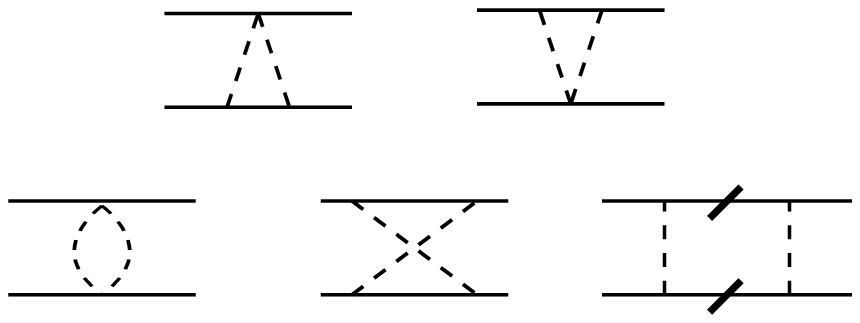}}
\smallskip
\caption{Fig.~1. One loop, 2-pion exchange Feynman graphs which contribute to
the 2-nucleon potential $V^{(2)}$.  The first four are 2-nucleon irreducible;
the last diagram is 2-nucleon reducible, and the poles from the slashed
propagators are not included in the $\int {\rm d}q_0$ loop integration. The
1-loop graphs corresponding to $\pi NN$ vertex renormalization  and pion wave
function renormalization are not pictured here, but enter at the same order (as
does nucleon wave function renormalization).}
\endinsert

A special comment must be made about the 1-nucleon potentials, $V^{(1)}$.
These diagrams consist solely of the 1-nucleon irreducible graphs. They include
both wave function renormalization (which begins at order $Q^2$) as well as
relativistic corrections to the nucleon propagator, which start at order
$Q^4/M^3$. The  structure of the latter terms is fixed by relativistic
invariance to reproduce the Taylor expansion of $\sqrt{\bfp^2+M^2}$.

A general $n$-nucleon Feynman diagram in the EFT  can be constructed by sewing
together the nucleon legs of $V^{(r)}$ potentials with $r\le n$;  one treats
the  $V^{(r)}$'s like vertices and the $\int {\rm d}q_0$ loop integrations pick
up the poles of all the connecting nucleon lines\foot{As pointed out by
Weinberg, this set of diagrams is more naturally described in the language of
time-ordered perturbation theory, but as there will be a mix of relativistic
pion propagators and nonrelativistic nucleon propagators, no formalism is
ideal, and we will keep to the language of covariant perturbation theory.}.

 The reason for this construction is that within the $V^{(r)}$ potentials,  all
nucleon propagators are off-shell and scale like $1/q_0\sim1/Q$. In contrast,
when one picks up the pole contribution from one of the  nucleon lines
connecting the $V^{(r)}$ ``vertices'', other nucleon lines will be almost
on-shell, and scale like $1/(Q^2/M)$.

Following Weinberg's arguments \weinberg, a contribution to the $r$-nucleon
potential $V^{(r)}$ with $\ell$ loops, $I_n$ nucleon propagators, $I_\pi$ pion
propagators, and $V_i$ vertices involving $n_i$ nucleon lines and $d_i$
derivatives,  scales like $Q^{\mu}$, where
\eqna\weinrel
$$\eqalignno{
\mu &= 4 l - I_n - 2I_\pi + \sum V_i d_i\ ,&\weinrel a\cr
\ell &= I_n + I_\pi - \sum V_i + 1\ ,&\weinrel b\cr
I_n+r&=\half\sum V_i n_i \ .&\weinrel c}$$
In this power counting we take $m_\pi\sim Q$ and treat factors of the $u$ and
$d$ quark masses at the vertices as order $Q^2$. Combining these relations
leads to the scaling law for the $r$-nucleon
potential $V^{(r)}$ ($r\ge 2$):
\eqn\wpo{\mu = 2+2\ell -r + \sum_i V_i (d_i+\half n_i-2)\ .}
Since chiral symmetry implies that the pion is derivatively coupled,  it
follows that $ (d_i+\half n_i-2)\ge 0$.  That implies that for a 2-nucleon
potential, $\mu\ge 0$, and that $\mu=0$ corresponds to tree diagrams.

It is straight forward to find the scaling property for a general Feynman
amplitude, by repeating the analysis that leads eq. \wpo,  treating the
$V^{(r)}$ potentials as $r$-nucleon vertices with $\mu$ derivatives, $\mu$
given by eq. \wpo.  However, while eq. \wpo\ was derived assuming that $\int
{\rm d}q_0\sim Q$ and nucleon propagators scaled like $\sim 1/Q$, we now take
them to scale like $Q^2/M$ and $1/(Q^2/M)$ respectively.    A general Feynman
diagram is constructed by stringing together $r$-nucleon potentials $V^{(r)}$.

For two nucleon scattering the situation is particularly simple, since the
diagrams are all ladder diagrams, with  $n$ insertions of  $V^{(2)}$'s acting
as ladder rungs.   Each loop of the ladder introduces a loop integration (${\rm
d}q_0{\rm d}^3\bfq\sim Q^5/M$) and two nucleon propagators ($\sim M^2/Q^4$) to
give a net factor of $(QM)$ per loop.
If we expand
$V^{(2)} = \sum_{\mu=0}^\infty  V^{(2)}_\mu $,
where $V_\mu^{(2)}\sim Q^\mu$,
then a  2-nucleon diagram whose $i^{th}$ rung is the generalized potential
$V^{(2)}_{\mu_i}$
scales as
\eqn\ourpo{Q^\nu (QM)^L\ ,\qquad
\nu = \sum_{i=1}^{L-1} \mu_i\ ,\qquad {\rm (2-body\ scattering)}}
where  $L$ is the number of loops (external to
the $V^{(2)}$'s).
With insertions of $V^{(1)}$ along the nucleon propagators, which serve as
relativistic corrections, there will be additional powers of $(Q^2/M^2)$;
likewise, an expansion of retardation effects in $V^{(2)}$ can be treated like
the nonrelativistic expansion.

Since $\mu_i\ge 0$, the leading behaviour of the 2-nucleon
amplitude is $(QM)^L$.  If one treats $M\simeq Q^0$, it
follows that perturbation theory is adequate for describing
the 2-nucleon system at low $Q$. In order to explain the
nonperturbative effects one sees (\eg, the deuteron, or the
large scattering length in the \iso\ channel) one must
conclude that $ M\simeq 1/Q$ in a consistent power counting
scheme.  Thus the effective field theory calculation must be
an expansion in $\nu$, given by eqs. \wpo, \ourpo.  To leading
order ($\nu=0$) one must sum up all ladder diagrams with
insertions of $V^{(2)}_0$ potentials with $\mu=0$.  At
subleading order one includes one insertion of  $V^{(2)}_1$
and all powers of $V^{(2)}_0$, etc.

\topinsert
\centerline{\epsfxsize=4.2in\epsfbox{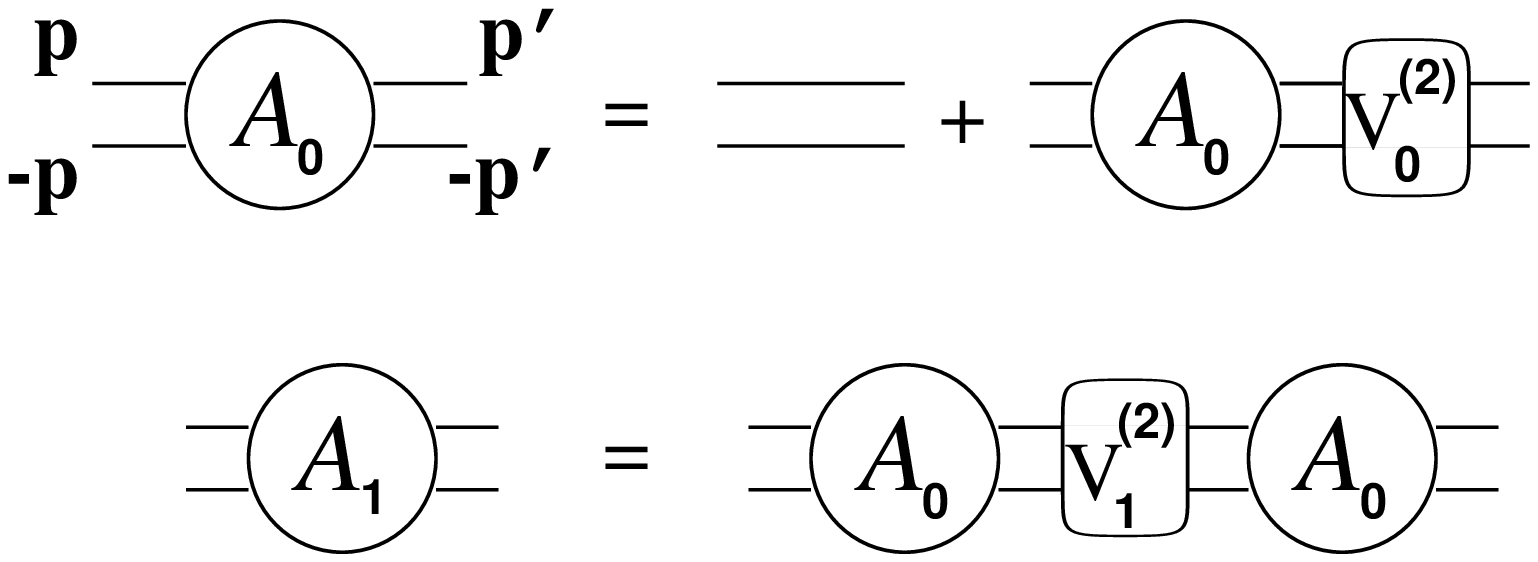}}
\smallskip
\caption{Fig. 2. The first two terms in the  EFT expansion for the Feynman
amplitude
($T$-matrix) for nucleon-nucleon scattering in the center of mass frame.  The
leading amplitude $\CA_0$
scales like $Q^0$ and consists of the sum of ladder diagrams with the leading
($\mu=0$) 2-nucleon potential $V^{(2)}_0$ at every rung;  the subleading
amplitude $\CA_1$ scales like $Q$; it consists one insertion of $V^{(1)}_1$
(1-loop nucleon wavefunction renormalization) or one insertion of the
subleading ($\mu=1$) 2-nucleon potential $V^{(2)}_1$, dressed by all powers of
the leading interaction $V^{(2)}_0$. }
\endinsert

The program advocated by Weinberg is to solve the Schr\"odinger equation
with the kernel $V^{(2)}$ expanded to a given order in  $\mu$.  An alternative
one
might consider is to expand the Feynman amplitude $\CA$ in powers of $\nu$;
this is
an equivalent procedure at $\nu=0$, but is unsatisfactory for higher $\nu$ as
the expansion violates unitarity.  We will argue in subsequent
sections that for systems with a large scattering length (e.g. $NN$ scattering)
the best procedure is to expand $\kcd = (4\pi/M){\rm Re}[1/\CA]$ in powers of
$\nu$,
where $\bfp$ is
the momentum in the center of mass frame, and $\delta$ is the phase shift. As
discussed below, this expansion preserves unitarity and is expected to converge
much faster than the expansion of the kernel $V^{(2)}$,
particularly in systems with a large scattering length.

A further modification we propose below is in the basic power counting scheme.
In particular, we will argue that there are explicit powers of $M$ in the
coefficients of operators of the EFT, which decreases their order in the EFT
expansion.  We return to this point below, once we have presented the
calculational techniques that lead to this result.

\subsec{Feynman diagrams and the Schr\"odinger equation}
Feynman diagrams are the usual tool
for computing perturbative amplitudes requiring renormalization, while the
Schr\"odinger equation is used to solve nonperturbative problems
in potential scattering.  As we will need to do both simultaneously,
we  briefly review here the connection between Feynman diagrams and
the Schr\"odinger equation.

Consider the integrals  arising from the ladder loops in the diagrams of
fig.~2:
\eqn\loopi{I=\igralf{q} V^{(2)}(p,q)
{i\over (E/2 + q_0 -\bfq^2/2M +i\epsilon)}
{i\over (E/2 - q_0 -\bfq^2/2M +i\epsilon)}
V^{(2)}(q,p')\ .}
In the above expression, $M$ is the nucleon mass, $E$ is the center of mass
kinetic energy. (As we will focus entirely on the 2-nucleon
problem for the rest of the paper, we will henceforth refer to the 2-nucleon
potential $V^{(2)}$ simply as $V$). Following the rules of the previous
section, the $\int {\rm d} q_0$ integral only picks up the pole contribution
from the nucleon propagators at $q_0=\pm(E/2-\bfq^2/2M)$.  Since $q_0\sim
Q^2/M$ one can consistently take $q_0\sim 0$ in $V$ (\ie, ignore retardation)
to the order we will be working.
In this approximation
\eqn\loopii{I\sim  i\igralt{q} V(\bfp,\bfq)
{1\over (E -\bfq^2/M +i\epsilon)}
V(\bfq,\bfp')\ .}
The connection
between the above expression and the Schr\"odinger equation is clarified by
defining the free retarded Schr\"odinger Green's function operator for the
2-nucleon system
\eqn\gogo{\hat G^0_E = {1\over (E-\hat H_0+i\epsilon)}\ ,\qquad \hat
H_0={\bfp^2\over M}\ ,}
Matrix elements of $\hat G_E^0$ and the potential operator, $\hat V$, between
momentum eigenstates are given by
\eqn\gvmats{
\bra{\bfp} \hat G^0_E\ket{\bfp'} = {(2\pi)^3\delta^3(\bfp-\bfp')\over(E
-\bfp^2/M
+i\epsilon)}\ ,\qquad \bra{\bfp} \hat V\ket{\bfp'} = V(\bfp,\bfp')
\ .}
The sum of ladder diagrams can then be expressed as
\eqn\lad{\eqalign{
i\CA &= -i\bra{\bfp}\(\hat V + \hat V\hat G^0_E\hat V + \hat V(\hat G^0_E\hat
V)^2+ ...\)\ket{\bfp'}\cr
&=-i\bra{\bfp}\hat V(1+\hat G_E\hat V)\ket{\bfp'}\cr
&=-i\bra{\bfp} (\hat G^0_E)^{-1}\hat G_E(\hat G^0_E)^{-1}\ket{\bfp'}\ ,} }
where $\hat G_E$ is the full Green's function with potential $\hat V$:
\eqn\ghdef{
\hat G_E = {1\over (E-\hat H+i\epsilon)}\ ,\qquad \hat H=\hat H_0+\hat V
\ .}

We can define the state
\eqn\chidef{\ket{\chi_{\bfp}} \equiv (1+\hat G_E\hat V)\ket{\bfp}=\hat
G_E(\hat G^0_E)^{-1}\ket{\bfp} }
with $\bfp^2/M=E$,
which is seen to be the exact scattering solution of interest:  it satisfies
the full Schr\"odinger equation since
\eqn\schroopa{
(\hat H-E)\ket{\chi_{\bfp}}=-\hat G_E^{-1}\ket{\chi_{\bfp}}=(\hat
H_0-E)\ket{\bfp}=0\ ,
}
and takes the appropriate asymptotic form for large $r$
\eqn\scattasym{
 \chi_{\bfp}({\bf r}) \to e^{i\bfp\cdot{\bf
r}} + {f(\theta,\phi)\over r} e^{i{\rm pr}}
\ ,}
since $\bra{r}\hat G_E\hat V\ket{r'}\propto 1/r$ for large $r$.
The Feynman
amplitude \lad\
can be expressed in terms of $\chi$ as
\eqn\laddi{i\CA = -i\int {\rm d}^3 r\, e^{-i\bfp\cdot{\bf r}}
V({\bf r}) \chi_{\bfp'}({\bf r})\ ,}
which is  $(-i)$ times the conventional expression for the $T$-matrix
(see, for example,
\ref\goldwat{M.L. Goldberger and K.M Watson, {\it Collision Theory},
John Wiley and Sons, Inc. (1964). }).
For $s$-wave scattering at center of mass
momentum ${\bf p}$, $\CA$ can be conveniently related to the phase shift
$\delta(\bfp)$  by the relation
\eqn\kcotd{\eqalign{
\kcd &= i|\bfp| + {4\pi\over M}{1\over \CA}\cr
&= -{1\over a} + \half r_0 \bfp^2 + \ldots\ ,}}
where $a$ is the scattering length and $r_0$ is the effective range.  For \iso\
($np$) scattering, these parameters are measured to be
(see  \ref\effr{M.A. Preston and R.K. Bhaduri, {\it Structure of the Nucleus},
Addison-Wesley Publishing Company, Reading , Massachusetts (1975);
2nd printing (1982).})
\eqn\aro{a= -23.714\pm 0.013 \ {\rm fm} \  ,
\qquad r_0 = 2.73\pm 0.03 \ {\rm fm}\ .}

The above discussion is complete when the potential $V$ is less singular than
$1/r^2$ at the origin, in which case the terms in the series of eq. \lad\ are
well defined.  However, in effective field theories, the potential will in
general have more singular behavior, such as $1/r^2$, $1/r^3$,  $\delta^3({\bf
r})$,
and worse. Such potentials do not allow a conventional solution to the
Schr\"odinger equation, or equivalently, lead to divergent diagrams in the
field
theory.  In the field theory it is well known how to deal with divergences ---
one merely regulates the integrals and then renormalizes the couplings of the
theory, absorbing terms that diverge as the cutoff is removed into the
definitions of the renormalized couplings.  When this is done, there is no
cutoff dependence in the theory.  In
this paper, we show how to sum up the leading diagrams using dimensional
regularization and the \ms\ subtraction scheme for the case of \iso\
nucleon-nucleon scattering.
This is equivalent to solving the dimensionally
regulated Schr\"odinger equation.
The advantages of our procedure are that dimensional
regularization with the $\msb$ scheme   preserves chiral
symmetry and simplifies computations.   Since the
renormalization scale $\mu$ introduced by $\msb$ (or any mass
independent scheme, such as $MS$) only enters in logarithms,
EFT power counting arguments are particularly transparent,
unlike when a momentum cutoff procedure is used.

\newsec{The effective theory with nucleons alone}

Although the power counting of the previous section assumed $Q\sim m_\pi$ and
explicitly included pion propagation, the analysis also applies to a lower $Q$
regime where the pion plays no role.  We analyze this case first as it is
analytically more accessible and quite instructive.

At very low energy $NN$ scattering we may consider an effective field theory
consisting solely of nucleon fields; all other degrees of freedom, such as
$\pi$'s, $\Delta$'s, $\rho$ and $\omega$ mesons  have been integrated out, and
their effects are subsumed in the coupling constants of the effective theory.
Note that this effective theory has nothing to do with chiral symmetry;  in
fact, it treats the pion as very heavy compared to momenta of interest, which
is the opposite of the chiral limit.

The EFT consists of all local nucleon interactions allowed by rotational
invariance,
isospin symmetry (which we assume to be exact in this paper) and parity.
For 2-nucleon
scattering, the only interactions that are of relevance are the operators with
four nucleon fields, as well as relativistic corrections to the nucleon
propagator;  we will be able to ignore the latter to the order we are working.
In such a theory, the only diagrams that contribute to the 2-body potential $V$
are tree diagrams.  It follows from eq. \wpo\ that each 4-nucleon operator has
a scaling dimension $\mu_i=d_i$, where $d_i$ is the number of derivatives in
the acting at the vertex.   Eq. \ourpo\ then tells us that the leading
contribution to the amplitude has $\nu=0$, and that $\CA_0$ is given by the
bubble sum of 4-nucleon operators with no derivatives.  At $\nu=2$, $\CA_2$ is
given by one insertion of a 2-derivative, 4-nucleon operator, dressed by the
no-derivative operator, as in fig.~2, etc.

The effective Lagrangian for this theory is given by
\eqn\lagi{\eqalign{
\CL &= N^{\dagger} i \partial_t N - N^{\dagger}{\nabla^2\over 2 M} N
-\half C_S (N^\dagger N)^2 -\half C_T (N^\dagger \vec\sigma N)^2 \cr &-
\frac{1}{4} C_2 \(N^\dagger {\nabla^2}N\)(N^\dagger N) + h.c.+\ldots}}
where  $\vec
\sigma$ are the Pauli matrices acting on spin indices, and the ellipses refer
to additional 4-nucleon operators involving two or more derivatives, as well as
relativistic corrections to the propagator.
The coefficients $C_S$ and $C_T$ of dimension ${\rm (mass)}^{-2}$ are the
couplings introduced by Weinberg
\weinberg; $C_2$ is a coupling of dimension ${\rm (mass)}^{-4}$. The  values of
$C_S$, $C_T$, $C_2$ are renormalization scheme dependent.

Nucleon scattering in the $^1S_0$ channel only depends on $C_S$ and $C_T$ in
the linear combination $C=(C_S-3C_T)$, and so the leading contribution to the
potential is
\eqn\vzero{V_0(\bfp,\bfp') =   C.}
Similarly, on can show that while there are a number of operators with four
nucleon fields and two derivatives, only the linear combination proportional to
$C_2$ in eq. \lagi\ contributes to $V_2$ in the \iso\ channel.  It will be
convenient for later discussion of the momentum expansion to define
\eqn\lamdef{C_2\equiv {C\over \Lambda^2}\ ,}
where $\Lambda$ is a parameter with  dimension of mass.   With this definition
the next to leading order contribution to the 2-nucleon potential is
\eqn\vtwo{V_2(\bfp,\bfpp)=  C\({\bfp^2+\bfpp^2\over 2 \Lambda^2}\)\
.}

\subsec{The $\nu=0$ calculation}

The ladder graphs in fig.~2 for the leading part of the amplitude $\CA_0$  can
be summed trivially with the kernel $V_0$ in eq. \vzero, since the expression
\lad\ is a geometric series.  Using dimensional regularization one finds
\eqn\isoamp{ i\CA_0 = {-iC \over 1- C {\tilde G}_E^0 ({\bf 0,0})}=
 {-iC \over 1+iC M|\bfp|/4\pi}\ .}
The quantity
\eqn\gzerol{ {\tilde G}_E^0 ({\bf 0,0}) = \int {d^n {\bf q}\over (2\pi)^n}
\int{d^n {\bf q}^\prime \over (2\pi)^n} \langle {\bf q}|{\hat G}_E^{(0)}|{\bf
q}'\rangle,}
is simply the coordinate space representation of the free Green's function
${\tilde G}_E^0({\bf r,r'}) = \bra{{\bf r}} 1/(E-\hat H_0+i\epsilon)\ket{{\bf
r'}}$ evaluated at ${\bf r}={\bf r}'=0$, with  reduced mass  $M/2$. This
corresponds to a
divergent one loop graph, which  in dimensional regularization is given by:
\eqn\gzero{\openup 2\jot\eqalign{
\tilde{G}_E^0 ({\bf 0,0})  &=
 \int {{\rm d}^{n} \bfq\over (2\pi)^{n}}\,{1\over (E -
\bfq^2/M+i\epsilon)}\cr
&= -\( 4 \pi\)^{-n/2} M (-ME-i\epsilon)^{(n-2)/2} \Gamma(1-n/2)\cr
&\too{n\to 3}{M\sqrt{-ME -i\epsilon}\over 4 \pi}=
{-iM|\bfp |\over 4\pi}\ .}}

Even though minimal subtraction introduces a renormalization scale
$\mu$, one finds that in dimensional regularization $\tilde{G}_E^0 ({\bf 0,0})$
is
finite as $n \rightarrow 3$ and so $C$ is independent of $\mu$, satisfying
the
trivial renormalization group (RG) equation
\eqn\rgi{\mu {\partial \ \over \partial \mu}\({1\over C}\) =0\ .}
The value of $C$ is determined by experiment via eq. \kcotd\
which fixes the threshold
amplitude to be $ \CA = - 4\pi a/M$, where $a$ is the scattering
length.
It follows from eqs. \isoamp\ and \gzero, using $M=940 {\rm MeV}$,
that
\eqn\czval{C = {4\pi a\over M}=-\({1\over 25\ \MeV}\)^2. }
The expression for the scattering amplitude \isoamp\ may be rewritten as
\eqn\ert{i\CA_0 = i{4\pi / M \over -1/a - i |\bfp|}\ ,}
which is recognized as the effective range theory expression for the amplitude
given a scattering length $a$ and effective range $r_0=0$:
\eqn\nuo{\kcd=-{1\over a}\ .}
  This is reasonable,
since the interaction \lagi\ is local.  In fact, it is shown in the appendix
how  \czval\ may be derived by solving the Schr\"odinger equation with a
potential $V( {\bf r})= C\delta^{(3)}({\bf r})$.

\subsec{The $\mu=2$ calculation}

When using an effective Lagrangian, it is important to know at what momentum it
fails. The momentum expansion in the effective Lagrangian \lagi\  breaks
down at the scale $Q\sim\Lambda$, where $\Lambda$ is the scale set by $V_2$
\vtwo.  To determine $\Lambda$, we must perform a second order calculation, at
$\nu=2$. How to do so is ambiguous:  if one expands the Feynman amplitude to
order $\nu=2$ as in fig.~2, one destroys unitarity.  In this section we will
follow Weinberg's prescription, namely to expand $V$ to second order (\ie,
$\mu=2$), and then
sum its effects on the amplitude to all order. Doing so includes the exact
expressions for the order $\nu=0$ and $\nu=2$ parts of the full amplitude, and
keeps parts of the higher order terms (from multiple insertions of $V_2$).  In
the following section we will consider an alternative calculation.

To next to leading order, the 2-nucleon potential $V$ is given by
\eqn\vertval{
V(\bfp,\bfp') = V_0+V_2 = C\(1+{\bfp^2+\bfp'^2\over
2\Lambda^2}\)\ .}
It is possible to sum all of the ladder diagrams with the vertex \vertval\ in
the \ms\ scheme;  as shown in appendix~B, one merely replaces $C$ in eq.
\isoamp\ by $C(1+\bfp^2/\Lambda^2)$:
\eqn\isoampii{
i\CA_{V_2} = {-i \over 1/[C(1+  \bfp^2/\Lambda^2 )]+
iM|\bfp|/4\pi  }\ ,}
where the subscript $V_2$ denotes that we have followed Weinberg's prescription
and expanded $V$ (rather than $\CA$) to subleading order $\mu=2$.
Since $E=\bfp^2/M$, the above expression for the amplitude can be expressed as
a prediction for $\kcd$ by means of eq. \kcotd:
\eqn\kcotdi{\kcd = -\({4\pi\over M}\){1\over C(1+\bfp^2/\Lambda^2)}\ .}
We can fit our two free parameters $C$ and $\Lambda^2$ to low energy
scattering data by expanding $\kcd$ to order $\bfp^2$ and fitting to the
measured scattering length and effective range \aro. The result is
\eqn\cvalsi{C = {4 \pi a\over M}=  -\({1\over 25\ {\rm MeV}}\)^2\
,\qquad\qquad {1\over \Lambda^2} = \half
r_0 a = -\({1\over 35\ {\rm MeV}}\)^2 \ .}
Since $\bfp\simeq \Lambda$ is the scale at which $V_0\simeq V_2$, we expect the
effective theory with nucleons alone to work well  at
center of mass momenta $|\bfp|\ll 35\ {\rm MeV}$, but to fail completely for
$|\bfp|\gtap 35\ {\rm MeV}$, corresponding to the lab
kinetic energy $T_{\rm lab}=2.6\ \MeV$.

\subsec{An alternative: expanding $\kcd$ to order $\nu=2$}

The result \cvalsi\ is very discouraging from the EFT point of view.  The
original premise in \S2 was that amplitudes could be expanded in powers of
$(Q/\Lambda)^\nu$, where $\Lambda$ was a mass scale typical of the particles
not included explicitly in the theory.  When pions are included, we would hope
that $\Lambda\sim m_\rho$; in the lower energy EFT we are considering here,
with the pion integrated out, one would expect $\Lambda\sim m_\pi$.  Instead,
eq. \cvalsi\ has $\Lambda\sim 1/\sqrt{a r_0}$; $r_0$ can be considered a
relatively short distance scale, but $a \sim -1/(8\ {\rm MeV})$ for the \iso\
channel,
which can hardly be called a typical QCD scale. In general, $a$ blows up as a
bound state (or nearly
bound state) approaches threshold.  Thus a small change in short distance
physics can make the EFT fail at arbitrarily low momenta.

The problem can be made more precise by examining the quantity $\kcd$.
Eqs. \kcotdi\  and \cvalsi\  imply that
\eqn\coti{\kcd = {1\over -a+\half a^2r_0\bfp^2}=-{1\over a}\sum_{n=0}^\infty
(-\half a r_0 \bfp^2)^n\ ,}
which has a radius of convergence at $\bfp^2\sim 1/(a r_0)$.  However, it is
known from general arguments that for a potential that falls of exponentially
as $e^{-m r}$ for large $r$, the true radius of convergence for $\kcd$ is
given by $\bfp^2\sim m^2$  \goldwat.
The quantity
$\kcd$ should have an expansion of the form
\eqn\cotii{\kcd \sim -{1\over a} +  \half r_0 \bfp^2\,\sum_{n=0}^\infty
(r_n^2\bfp^2)^n\ ,}
where the scales $r_n$ are typical of the range of the potential, $r_0\sim
r_n\sim1/m$.  None of the $r_n$'s are expected to diverge as $|a|\to \infty$.

How are we to reconcile eq. \cotii\ with our discovery in eq. \coti\ that the
scale of momentum variation in the EFT is set by the length scale
\hbox{$\sqrt{ar_0}$\ ?}\  The only possible answer is that the higher
derivative
operators in the EFT, although controlled by a scale that diverges as $|a|\to
\infty$, are actually highly correlated, and the effects that diverge with $a$
cancel.
To see how this works in the present theory, consider a
different  expansion than the one performed above: instead of expanding
$V$ to order $\mu=2$ and solving for the amplitude,  we will expand $\kcd$ to
order $\nu=2$.  In terms of a $\nu$ expansion of the Feynman amplitude (as in
fig.~2)
\eqn\asum{
\CA\  =\ \CA_0\  +\  \CA_1\  +\  \CA_2\  +\  \ldots
\ ,}
the expansion of $\kcd$ is given by
\eqn\kcotdii{\eqalign{\kcd &= i|\bfp| + {4\pi\over M}{1\over \CA}\cr
&= i|\bfp| + {4\pi\over M}{1\over \CA_0}\[1-\({\CA_1\over
\CA_0}\) + \({\CA_1^2-\CA_0\CA_2\over \CA_0^2}\)+\ldots\]\ .}}
 Note that to compute $\kcd$ to order $\nu_0$, one needs to compute $\CA_\nu$
only for $\nu\le \nu_0$, which involves perturbation theory in all but the
$\nu=0$ potential \foot{In a theory with just nucleons, all $\CA_\nu$ vanish
for odd $\nu$;  this is no longer true when pions are included.}.  In the
present EFT, we have found
\eqn\aexpzero{
\CA_0 = {C\over 1+i CM|\bfp|/4\pi} = {4\pi/M\over -1/a-i|\bfp|}}
(from eqs. \isoamp, \cvalsi), $\CA_1=0$, and
\eqn\aexptwo{\openup 2 \jot \eqalign{
\CA_2&=  -\({C\over 1+i CM|\bfp|/4\pi}\)^2 \({\bfp^2\over C \Lambda^2}\)
\cr &= -\({M\CA_0^2\over 4 \pi}\) \(\half r_0 \bfp^2\)\ ,}}
obtained by expanding eq. \isoampii\ to first order in $1/\Lambda^2$ and
substituting the values \cvalsi.  Substituting the above expressions into eq.
\kcotdii, we find that the $\nu=2$ expansion of $\kcd$ exactly reproduces
effective range theory,
\eqn\cotexp{\kcd = -{1\over a} + \half r_0 \bfp^2\qquad (\nu=2)\ .}

  In fact, with no long range interactions, effective range theory had to be
equivalent to the $\nu=2$ EFT expansion of $\kcd$, by dimensional analysis.
However, when pions are included the EFT expansion of $\kcd$ is  {\it not}
equivalent to effective range theory, since each order in the $\nu$ expansion
generates a complicated dependence for $\kcd$ on $\bfp$ and $m_\pi$ of order
$Q^\nu$.  In fact, as we show in the next section where we include pions, the
lowest order, $\nu=0$ calculation (with the scattering length $a$ as
experimental input) allows us to predict a nonzero value for the \iso\
effective range $r_0$.

\subsec{Comparison with data}

\nref\nij{Partial wave analysis of the Nijmegen University theoretical high
energy physics group, obtained from WorldWide Web page
http://nn-online.sci.kun.nl/}\topinsert
\centerline{\epsfxsize=3in\epsfbox{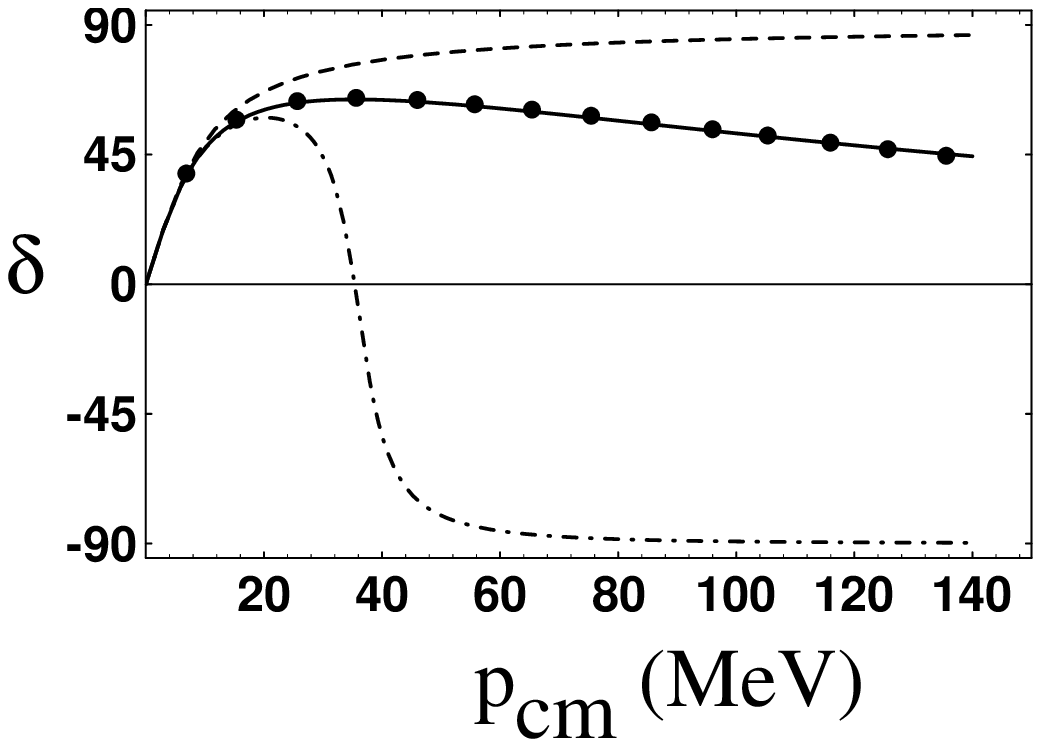}}
\smallskip
\caption{Fig. 3. \iso\ $np$ phase shifts in degrees plotted versus
center of mass momentum.   The dots are the \iso\ phase shift data from the
Nijmegen partial wave analysis \nij; the dashed, dash-dot and solid lines are
EFT calculations in a theory without pions.   The dashed line is the  $\nu=0$
result  from eq. \nuo; the dash-dot line is the  EFT result when the potential
is expanded to order $\nu=2$, eq. \coti;  the solid line (which lies along the
dots) is the EFT result when $\kcd$ is expanded to order $\nu=2$, eq.
 \cotexp.}
\endinsert
In fig.~3 we show a plot of the \iso\ $NN$ phase shift produced by the Nijmegen
partial wave analysis \nij, compared with the three EFT analyses we have
performed without pions:
the $\nu=0$ calculation \nuo; the $\mu=2$ expansion of the kernel
$V$ \coti; and the $\nu=2$ expansion of $\kcd$ \cotexp.  The first two
calculations yield
results that agree well with data for $|\bfp|\ltap 20\ \MeV$, but differ
wildly above that scale.  This is what one would expect from the size of $C$
and $\Lambda^2$ in eqs. \czval, \cvalsi.  In contrast, the $\nu=2$ expansion of
$\kcd$ yields a result indistinguishable from the Nijmegen analysis beyond
$|\bfp|=m_\pi$. This demonstrates  the
nearly complete cancellation between operators  of different index $\mu$
discussed above, at momenta greater than $\Lambda$.

We do not wish to belabor this phenomenological success with this simplistic
model --- we have only shown that low energy EFT can
reproduce effective range theory, and we have only considered a single partial
wave.  However, we have demonstrated that expansion of $\kcd$ extends the range
of validity of the EFT beyond the scale set by the derivative
expansion\foot{This suggests that the EFT might be profitably formulated with a
light degree of freedom in the $s$-channel.}. 
There is also an important practical reason for preferring to
expand $\kcd$:  that is that the effects of $\mu>0$ interactions need only be
computed in perturbation theory, following the expansion \kcotdii.  This in
general leads
to a great simplification of the calculation.
Furthermore, it provides a
way to implement a consistent renormalization procedure, as we will
see in the next section, where we introduce pions.

\subsec{Rethinking the EFT expansion}

We showed above that the  effective Lagrangian is a momentum expansion in
powers of $ p^2 a r_0$, where $r_0$ is a typical hadronic scale (\ie $\sim 1$
fermi), while $a$ is the scattering length. This led us to the conclusion that
the object with a sensible momentum expansion is not the potential, but rather
${\rm Re}[1/\CA]$, the real part of the inverse Feynman amplitude.  Another
conclusion we can draw is that the power counting scheme presented in \S2 needs
modification --- eq.
\wpo\ in particular.  The leading, 4-fermion operator has a coefficient $4\pi
a/M$, and is treated as a $\mu=0$ contribution to the potential.  Since the
combination $MQ$ is assumed to be of degree $\nu=0$, it follows that powers of
$aQ$ appearing in the amplitude are also of degree $\nu=0$.  As the
effective Lagrangian is an expansion in $\sim\bfp^2 a r_0$, a $2d$ derivative
vertex does not contribute $\mu=2d$ to a graph, as assumed in eq. \wpo, but
rather $\mu=d$ as it scales like $(a Q^2)^d$.  This does not matter in the
present theory;  the ``$\nu=2$ calculation'' needs only to be renamed the
``$\nu=1$ calculation''.  However as we show below, there are striking
implications when pions are included in the EFT.

\newsec{The effective theory with nucleons and pions}

In order to extend the energy range over which the effective theory is useful,
it is necessary to include more light degrees of freedom.  The obvious
candidate to add to the theory is the pion.  In the previous section, analyzing
$np$ scattering in an effective theory without pions, we found that the
derivative expansion in the EFT broke down at a scale $\Lambda\simeq 35\ \MeV$.
 A sign that we are improving the utility of the EFT by including pions will be
whether or
not the scale set by the contact interactions becomes significantly higher.  As
we will show, that is the case.

Chiral symmetry mandates that pions couple to nucleons derivatively, or
proportional to powers of the quark masses.  In the power counting arguments of
\S2, we assumed $Q^2\sim m_\pi^2\propto m_q$, where $m_q$ are the $u$ and $d$
quark masses.  To determine which operators to include at a given order in the
EFT expansion, it is necessary to look to eqs. \wpo\ and \ourpo.  To compute
the 2-nucleon potential at order $\mu=0$ we include all tree level interactions
for which $\sum V_i(d_i+\half n_i-2)=0$.   That includes the 4-nucleon
interaction without derivatives, as well as 1-pion exchange with the
1-derivative axial vector coupling at each vertex.  
In this section we perform the leading ($\nu=0$) and subleading ($\nu=1$)
calculations exactly.

\subsec{The $\nu=0$ amplitude}

The $\mu=0$,  2-nucleon potential $V_0$ for $NN$ scattering in the \iso\
channel
is given  to leading order by one pion
exchange, plus a contact term:
\eqn\ope{
V(\bfp,\bfp') = C \ - \ \({g_A^2\over
2f_\pi^2}\){(\bfq\cdot\sigma_1\bfq\cdot\sigma_2)(\tau_1\cdot\tau_2)\over
(\bfq^2+m_\pi^2)\ }\ ,}
with $\bfq\equiv (\bfp-\bfpp)$.
The coupling $g_A =1.25$ is the axial coupling
constant,
$m_\pi = 140 {\rm MeV}$ is the pion mass,
and $f_\pi$ is the pion decay constant normalized to be
\eqn\fpidef{
f_\pi = 132\ {\rm MeV}
 \ ,}
compared to other common normalizations $f_\pi = \sqrt{2} (93\ \MeV) = (186\
\MeV)/\sqrt{2}$.  As in the previous section, $C$ is a free parameter which
will be computed in $\msb$ subject to the  condition that we correctly
reproduce the measured threshold scattering amplitude (\ie, the scattering
length $a$).
Since we are exclusively interested in the \iso\ channel ($I=1$)  we can
express $V_0$ as
\eqn\opei{ V_0 ({\bf p,p'})=\tilde C + V_\pi(\bf p,\bf p')\ ,}
 where
\eqn\cydef{\tilde C\equiv \(C+ {g_A^2 \over 2 f_\pi^2}\)\ ,\qquad
V_\pi(\bfp,\bfp')\equiv -{ 4\pi\alpha_\pi\over (\bfq^2+m_\pi^2)}\ ,\qquad
\alpha_\pi\equiv \({g_A^2
m_\pi^2\over 8\pi f_\pi^2}\)\ .}
Note that while $V_\pi$ is the conventional one-pion
exchange (OPE) potential, our calculation will differ significantly from OPE
due to the $\tilde C$ contact interaction.  The contact term  includes not only
the $\delta^3({\bf r})$ contribution from one pion exchange,
but also the leading contribution in the derivative expansion of all shorter
distance effects, such as
2-pion exchange, intermediate $\Delta$'s, $\omega$ exchange, etc.

\topinsert
\centerline{\epsfxsize=4.5in\epsfbox{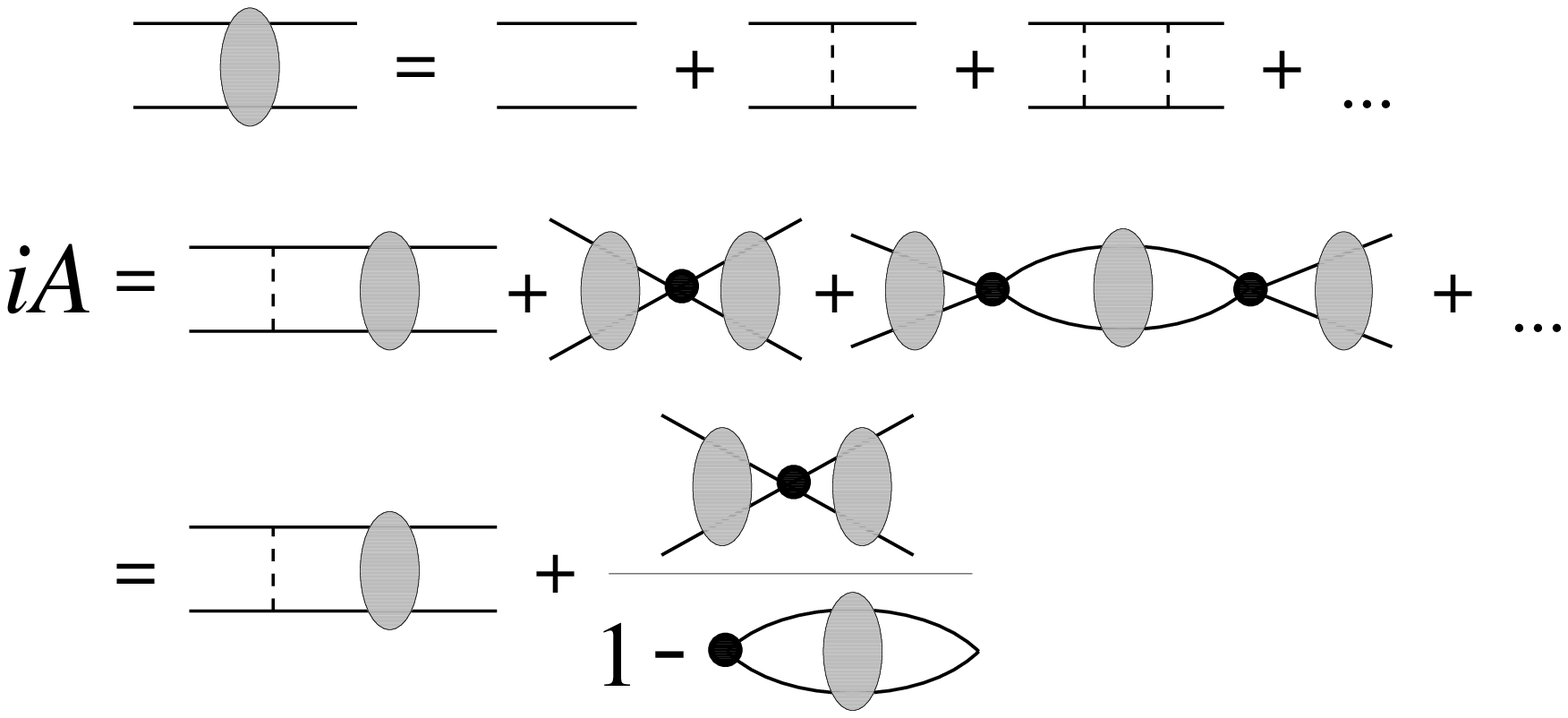}}
\smallskip
\caption{Fig. 4. Ladder diagrams for the $\nu=0$ contribution to the Feynman
amplitude $\CA_0$ are formally resummed by expressing the kernel
$V_0$ as a sum of a contact interaction proportional to $\tilde C$  and a
nonlocal interaction $V_\pi$
as in eq. \opei.  The shaded blobs consist of the ladder sum of $V_\pi$
interactions (dashed lines), while the black vertices correspond to a factor of
$\tilde C$.
}
\endinsert
It is not possible to compute the ladder sum with the above kernel
analytically, but we are able to express it in terms of several quantities that
can be computed numerically with ease.
Most importantly, we are able to renormalize the nonperturbative
amplitude analytically.
To achieve this,
the ladder diagrams  are formally summed as in fig.~4 to yield   the Feynman
amplitude\foot{Here we give a Feynman diagram
approach, while in appendix A we show how the Schr\"odinger equation
corresponding to the kernel \ope\ can be solved directly.}
\eqn\famp{i\CA_0 = i\CA_\pi -i{\tilde C \[ \chi_{\bfp}({\bf 0})\]^2\over 1 -
\tilde
C
{\tilde G}_E({\bf 0,0})}\ ,}
where the quantities $\CA_\pi$, $\chi_{\bfp}({\bf 0})$ and $\tilde{G}_E ({\bf
0,0})$
are
the sub-diagrams pictured in fig.~5.
\topinsert
\centerline{\epsfxsize=4.5in\epsfbox{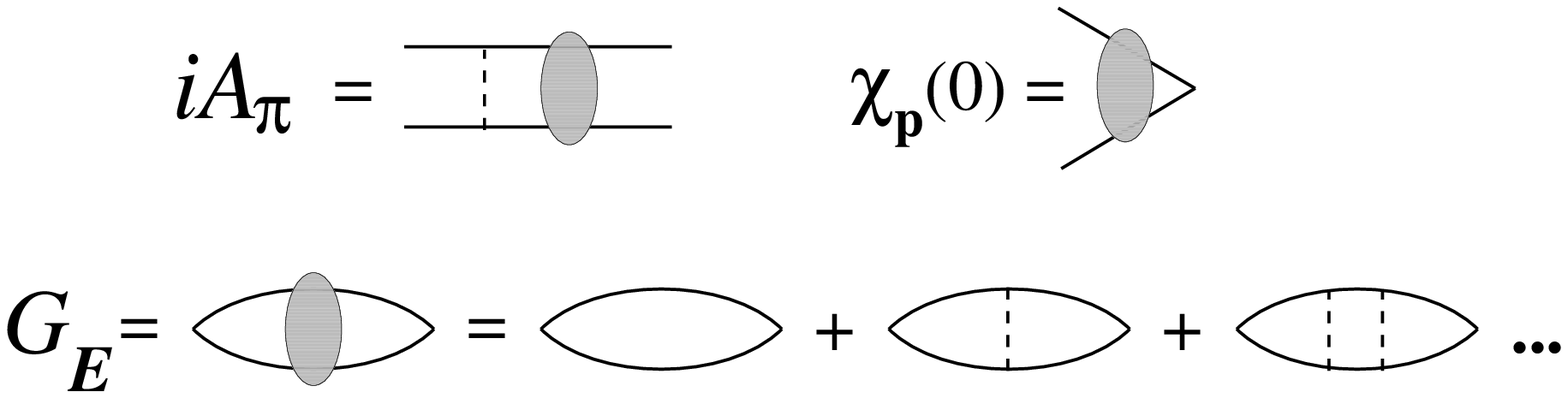}}
\smallskip
\caption{Fig. 5. Subdiagrams defining the quantities $\CA_\pi$,
$\chi_{\bfp}({\bf 0})$ and $\tilde{G}_E ({\bf 0,0})$ used in eq. \famp.
$\CA_\pi$
and $\chi_{\bfp}({\bf 0})$
are finite, as are all but the first two diagrams in the expansion of
$\tilde{G}_E ({\bf 0,0})$.
The shaded blob is defined diagrammatically in fig.~4. Dashed lines are
insertions of $V_\pi$, eq. \cydef}
\endinsert

The quantity $\CA_\pi$ is just the amplitude one finds in the pure Yukawa
theory with potential $\hat V_\pi$, i.e, the usual OPE result:
\eqn\icy{\eqalign{
i\CA_\pi&= \bra{\bfp}\hat V_\pi (1+ \hat G_E \hat V_\pi)\ket{\bfp'}\ ,\cr
\hat G_E &= {1\over E- \hat H_0- \hat V_\pi+i\epsilon}\ ,}}
while
\eqn\chipdef{
\chi_{\bfp}({\bf 0}) = \igralt{q} \bra{\bfq}(1+\hat G_E\hat V_\pi)\ket{\bfp}
}
is the OPE wave function at the origin. Both $\CA_\pi$ and $\chi_{\bfp}({\bf
0})$ can
be computed numerically by solving the Schr\"odinger equation with the Yukawa
potential $V_\pi$.  This is discussed in appendix A, where the solutions are
plotted (figs.~7,8).

The quantity ${\tilde G}_E ({\bf 0,0})$ is the coordinate-space propagator from
the
origin to the origin in the presence of $V_\pi$; it is divergent but can be
defined in \ms
\eqn\GEint{
{\tilde G}_E ({\bf 0,0}) = \dint{{\rm d}^3\bfq\over (2\pi)^3}{{\rm
d}^3\bfq'\over (2\pi)^3} \bra{\bfq'}\hat G_E\ket{\bfq}
 \ .}
Divergences occur in only the first two graphs in the perturbative
expansion for ${\tilde G}_E ({\bf 0,0})$ shown in fig~5.
These two graphs can be computed in \ms, while the remaining graphs
can be summed by numerically computing the propagator
$\tilde{G}_E({\bf r},{\bf 0})$; see
appendix~A for details.
As a result of renormalization, ${\tilde G}_E ({\bf 0,0})$
is replaced by the finite ${\tilde G}^{\msb}_{E} ({\bf 0,0})$, while
the bare $\tilde C$ is replaced by the renormalized  ${\tilde C}_{\msb}(\mu)$
which is to be fit to experiment.  Note that both quantities now depend on a
renormalization scale $\mu$;  however the amplitude $\CA_0$ is $\mu$
independent.   We can compute  the renormalization group equation for
${\tilde C}_{\msb}(\mu)$, which is given by
\eqn\rgeq{\mu{\partial\ \over \partial\mu}{1\over {\tilde C}_{\bar{MS}}(\mu) }=
-{\alpha_\pi
M^2\over 4\pi}\qquad\qquad\qquad{(\bar{MS})}\ .}
This result is derived in appendix A (see eq. (A.28)).
Throughout this paper we will be quoting
values for coupling constants renormalized at the scale $\mu=m_\pi$;  the
reason for this is that loop diagrams omitted at a given level of the $\nu$
expansion  bring in factors of $\ln m_\pi^2/\mu^2$, and so
choosing $\mu\sim m_\pi$ is expected to optimize the perturbation expansion for
$|\bfp|\ltap m_\pi$.   Note that for  $\tilde C_{\msb}(\mu)$ negative,
$|1/\tilde C_{\msb}(\mu)|$ increases with increasing renormalization scale
$\mu$.

After solving for $\CA_\pi$, $\chi_{\bfp}$ and $G_E^{{\bar {MS}}}$ numerically,
one can compute the amplitude \famp\ and fit $C_{\bar{MS}}(\mu)$ in order to
obtain the correct scattering length.    We find
\eqn\newcval{
C_{\bar{MS}}(\mu)\biggl.\biggr|_{\mu=m_\pi} = -\({1\over 79\ \MeV}\)^2
\quad (\nu=0)
 \ ,}
($C_\msb \equiv \tilde C_\msb-g_A^2/2f_\pi^2$)
which shows a substantial improvement over the value  $C=-(1/25\ \MeV)^2$
obtained in the pure nucleon effective theory at $\nu=0$,  eq. \czval.
Furthermore, once $C_{\bar{MS}}$ is fixed to give the
correct scattering length, one can compute the effective range, and we find
\eqn\effpred{
r_0= 1.3\ {\rm fm}\qquad (\nu=0)
 \ ,}
which shows that this simple effective theory with a contact term and one-pion
exchange can account for about half of the measured effective range, $r_0=2.7$
fm.  In fig.~7 we plot the phase shift determined from the amplitude $\CA_0$ in
eq. \famp\ as a
function of the center of mass momentum $|\bfp|$.

\topinsert
\centerline{\epsfxsize=4in\epsfbox{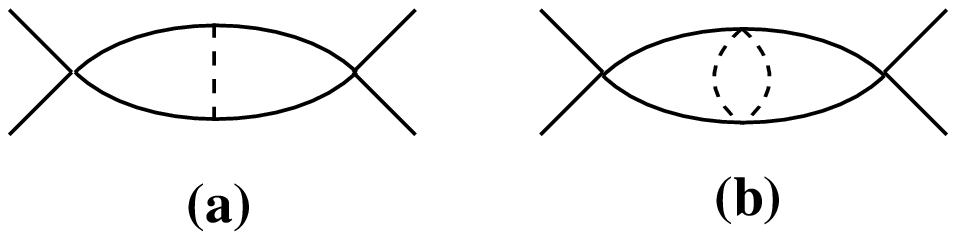}}
\smallskip
\caption{Fig.~6. Examples of graphs  with logarithmic divergences: (a) one
proportional to $M^2 m_\pi^2$; (b) another with a term  proportional to $M^2
\bfp^4$. The solid lines are nucleons, while the dashed lines are pions.}
\endinsert
\subsec{A new power counting scheme}

In section \S3.5  we pointed at that the parametrically large coefficients of
operators in the effective Lagrangian forced us to revise our power counting
scheme; however the revised counting had little practical effect in the  theory
without pions. However, with pions the story is different.  In particular,
equation \rgeq\ above mandates that we modify our EFT expansion scheme
significantly.  The logarithmically divergent graph that leads to the scaling
of $C$  is shown in fig.~6(a); in $4-2\epsilon$ dimensions it gives rise to the
pole
\eqn\pole{-{1\over \epsilon} {\alpha_\pi M^2\over 16 \pi} =
-{1\over \epsilon} {g_A^2 m_\pi^2 M^2\over 128\pi^2 f_\pi^2}\ .}
By our power counting, it is perfectly consistent to find such a divergence at
$\nu=0$, since $m_\pi^2 M^2$ has degree $\nu=0$.  
However, note that the counterterm required to absorb this divergence is 
of the form $ (N^\dagger \CM_q N)(N^\dagger N)$, where
$\CM_q$ is the quark mass matrix.  In the power counting scheme described in
\S2, such an operator, which is higher order in a chiral expansion, was
considered to be a $\mu=2$ operator, since it was assumed that the coefficient
of the operator was set by a ``typical'' QCD scale.  Instead, we see from eq.
\pole\ that the coefficient of this operator has an explicit factor of $M^2$,
and so is actually of degree $\mu=0$~\foot{The connection between the
coefficient of a $1/\epsilon$ pole, and the natural size of the coefficients of
operators in the effective Lagrangian is ``naive dimensional analysis''.  The
idea is simply that, because of the RG flow such as in eq. \rgeq, even if the
coefficient $C(\mu)$ of an operator is small for some reason at a scale
$\mu_0$, at a scale $\mu = \mu_0\times \CO(1)$, $C(\mu)$ will have flowed to
the magnitude of the coefficient of the $1/ \epsilon$ pole, which is
therefore considered its ``natural'' size.}.

Evidently the EFT expansion is {\it not} equivalent to a chiral expansion in
$m_\pi$, since a calculation at finite  order in the EFT expansion can be
modified by arbitrary powers of $M^2m_\pi^2$.
This may sound like a disaster for the EFT expansion, but it  actually is not.
At $\nu=0$, for example, we still have only one contact interaction $C$, and
whether or not this contains contributions from all orders in $m_\pi$ has no
effect on the predictive power of the calculation.

What would be a disaster is if there were counterterms needed proportional to
powers of $M^2 \bfp^2$.  If that were the case, the theory would not be
predictive, as an entire form factor would be needed to describe scattering at
lowest order in the EFT expansion.  However such terms do not arise.  
Consider
the diagram fig.~6(b), which contributes to the $\nu=2$ calculation of the
amplitude.  The graph  is proportional to $C^2M^2/f^4\sim 1/({\rm mass})^6$.
Since the graph must be proportional to $ 1/({\rm mass})^2$, there can be a
logarithmic divergence (and hence a $1/\epsilon$ pole that requires a
counterterm) proportional to $C^2 M^2 \bfp^4/f^4$.  This is consistent with the
graph being $\nu=2$; it is also consistent with the power counting of the
theory without pions, where the $\bfp^4$ interaction was multiplied by $(a
r_0)^2$.  However, this result is not consistent  with the naive power counting
in \S2 that assigned degree $\mu=4$ to a four nucleon contact interaction
proportional to $\bfp^4$.

It is possible to make general power counting arguments that the four nucleon
contact terms in the  effective Lagrangian involve an expansion in $M^2
m_\pi^2$ and $M \bfp^2$.  Equation \wpo\ is therefore modified so that the
degree $\mu$ of a vertex gets a contribution $\Delta\mu = 1$ rather than
$\Delta\mu = 2$ from each factor of nucleon momentum squared, $\bfp^2$.
Furthermore, powers of $\CM_q$ in four nucleon contact interactions are assumed
to be accompanied by $M^2$ and do not increase $\mu$.

\subsec{The full $\nu=1$ amplitude with one-pion exchange}

{}From the above discussion we see that the $\bfp^2$ contact interaction
 --- given that its coefficient scales  $\sim M$ ---
is the
only operator in the effective Lagrangian  of degree $\mu=1$.  
Two pion exchange --- considered of equal order in the incorrect
power counting of \S2 --- contributes at order $\mu=2$, as does the $\bfp^4$
contact interaction. A full $\nu=1$ treatment of the \iso\ scattering amplitude
is therefore simple to obtain, while extending the analysis to $\nu=2$ is quite
ambitious.

As in \S3, we first consider the Weinberg expansion, summing up the $\mu=1$
potential $V=V_0+V_1$ to all orders, where $V_0$ is given in eq. \opei\ and
\eqn\vii{
V_1 = \tilde C\, \left( {\bfp^2+{\bfp'}^2\over 2 \Lambda^2}\right)
 \ .}
The resultant amplitude is (see appendix~B for details)
\eqn\fampii{
i\CA_{V_1} = i\CA_\pi -i{ \[\chi_{\bfp}({\bf 0})\]^2\over \[\tilde
C\(1-{\alpha_\pi m_\pi M/ \Lambda^2} + {\bfp^2/ \Lambda^2}\)\]^{-1}  -
{\tilde G}_E ({\bf 0,0})}
 \ ,}
where $\CA_\pi$, $\chi_\bfp({\bf 0})$ and $\tilde G_E({\bf 0,0})$ are defined
as in
the previous section, fig.~5.
Expanding the denominator of eq. (4.7) in powers of $1/\Lambda^2$ gives
\eqn\denom{
{1\over \tilde C}
\ +\  {1\over{\tilde C}} \left({\alpha_\pi m_\pi M\over \Lambda^2}
\ +\  {\bfp^2\over \Lambda^2}\right) \ +\  \ldots \ -\  {\tilde G}_E ({\bf
0,0})
 \  .}
The term ${\tilde G}_E$ is divergent and if it is defined using dimensional
regularization it has  an energy independent $1/\epsilon$ singularity.
Using the $\bar{MS}$ subtraction scheme, this divergence is
absorbed into a renormalization of
${\tilde C}$ changing it to ${\tilde C}_{\overline{MS}} (\mu)$.
Then in the
second term proportional to $1/\Lambda^2$ we must introduce a renormalized
$\Lambda_{\bar{MS}} (\mu)$ defined by ${\tilde C} \Lambda^2 = {\tilde
C}_{\overline{MS}} (\mu) \Lambda_{\overline{MS}} (\mu)^2$.
However, with ${\tilde C}$
and $\Lambda$ renormalized in this way there is no freedom to express the
higher order terms represented by the ellipses in eq. \denom\  in terms of
renormalized parameters.  This problem arises because we have not included
operators with more than two derivatives.  They are needed as counter terms to
render multiple insertions of  the two derivative operator in eq. \lagi\
finite.  This is equivalent to saying that no redefinition of couplings in eq.
\fampii\ can absorb the energy independent $1/\epsilon$ pole  in $\tilde
G_E({\bf 0,0})$.

A   procedure which we can follow, consistent to order $\nu=1$ in an expansion
of the amplitude, is to include all the higher derivative operators, absorb the
$1/\epsilon$, and then arbitrarily set the renormalized coefficients of the
higher order terms to zero.
This {\it ad hoc}  procedure results in the analog of eq.
\isoampii, with one pion exchange effects included:
\eqn\isoampia{
 i{\cal A}_{V_1} = i{\cal A}_\pi -i { [\chi_\bfp({\bf 0})]^2\over
\left[\tilde C_{\overline{MS}} \left(1 - \alpha_\pi m_\pi M /
\Lambda_{\overline{MS}}^2 + \bfp^2 / \Lambda_{\overline{MS}}^2 \right)
\right]^{-1} - \tilde G_E^{\bar{MS}} ({\bf 0,0})}
 \  .}
 A (numerical) fit to the measured scattering
length and effective range with the amplitude in eq. \isoampia\ gives
\eqn\rela{
C_{\overline{MS}} (\mu)\bigg|_{\mu = m_{\pi}} = -{1\over (125\ \MeV)^2}
 \   ,
\qquad\qquad
{1\over \Lambda_{\overline{MS}}^2(\mu)} \bigg|_{\mu = m_{\pi}} =
-{1\over (43\ \MeV)^2}
  \  .}

Comparing the above values with the analogue without pions, eq. \cvalsi, we see
that the inclusion of pions has greatly reduced the value of $C$, while not
significantly altering the scale $\Lambda$ of the derivative expansion.
Therefore, the range of utility of the EFT apparently remains disappointingly
small.  In particular, the scale $\Lambda$ is far below the Fermi momentum in
nuclear matter ($p_F\sim 280\ \MeV$) so that the theory would appear to be of
little utility in understanding nuclear physics.  Nevertheless, following the
discussion in \S 3.3, we expect an expansion of $\kcd$ to order $\nu=1$ will
work much better; this is indeed the case, as demonstrated in fig.~7.
\eqnn\azval\
\eqnn\atval\
\eqnn\kcdexpii\
\topinsert
\centerline{\epsfxsize=3in\epsfbox{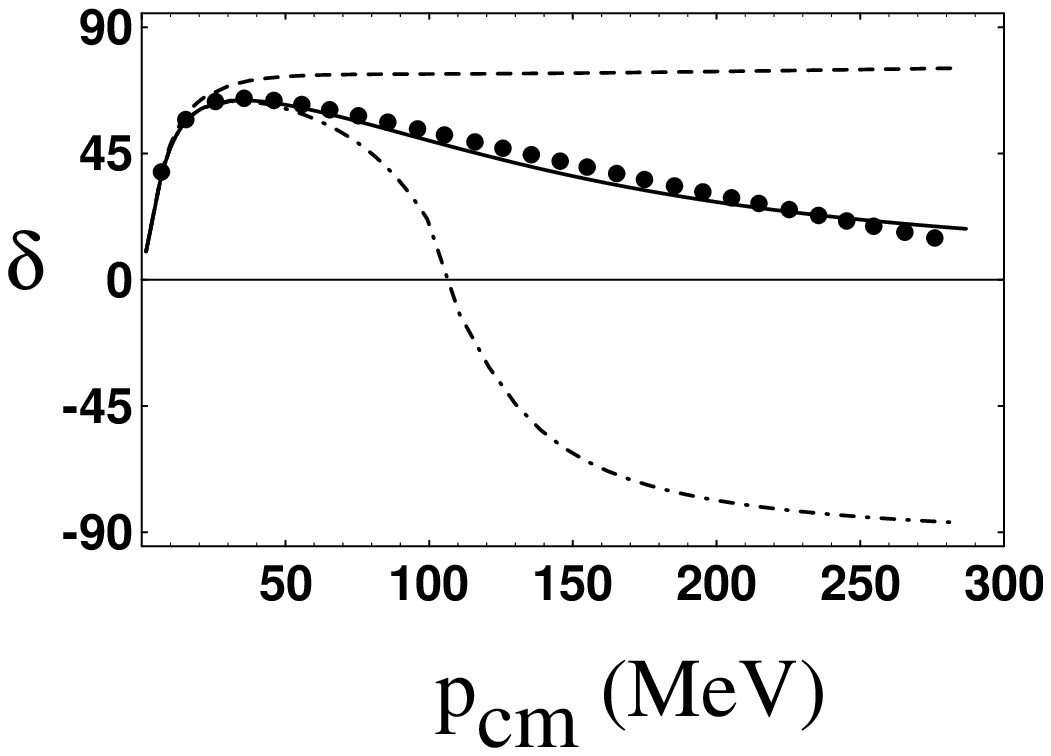}}
\smallskip
\caption{Fig. 7.  \iso\ $np$ phase shifts in degrees plotted versus
center of mass momentum.   The dots are the \iso\ phase shift data from the
Nijmegen partial wave analysis \nij; the dashed, dash-dot and solid lines are
EFT calculations in a theory with one pion exchange.   The dashed line is the
$\nu=0$
result  from eq. \famp; the dash-dot line is the  EFT result when the potential
is expanded to order $\nu=1$, eq. \fampii;  the solid line (which lies along
the
dots) is the EFT result when $\kcd$ is expanded to order $\nu=1$, eqs.
\azval---\kcdexpii.
Note that the momentum range of the plot extends to $2m_\pi$,
twice the range of fig.~3.}
\endinsert

To compute the $\nu=1$ expansion of $\kcd$  we need to
know $\CA_0$ and $\CA_1$.  The amplitude $\CA_0$ was computed in the previous
section in eq. \famp, with the substitution of the $\msb$ values for $\tilde C$
and $\tilde G_E$:
$$\CA_0 =
\CA_\pi - {[\chi_{\bfp}({\bf 0})]^2\over
1/\tilde C_{\msb}- \tilde G^{\msb}_E({\bf 0,0})}\ .\eqno\azval$$
The second order amplitude $\CA_1$ corresponds to the sum of graphs
indicated in fig.~2;  their sum is given by our previous calculation
\isoampia\ expanded to first order in $1/\Lambda^2$:
$$
\CA_1 = {[\chi_{\bfp}({\bf 0})]^2 \over
\tilde C_{\msb} \[1/\tilde C_{\msb}-\tilde G^{\msb}_E({\bf 0,0})\]^2}
\({\alpha_\pi m_\pi M/ \Lambda_\msb^2} - {\bfp^2/\Lambda_\msb^2}\) \
.\eqno\atval$$
The $\nu=1$ expansion of $\kcd$ is given by eq. \kcotdii:
$$\kcd = i|\bfp| + {4\pi\over M}{1\over \CA_0}\[1-\({\CA_1\over
\CA_0}\)\]\qquad (\nu=1)\ .\eqno\kcdexpii$$
This procedure is well defined from the point of view of renormalization:
Note that aside from the explicit factor of $\bfp^2$ in eq. \atval, there is
also complicated momentum dependence in $\CA_\pi$,  $\chi_\bfp({\bf 0})$ and
$\tilde G_E^\msb$ (OPE Feynman amplitude,  the OPE wave function at the origin,
 the renormalized OPE Green function at the origin respectively).
Thus the  terms in the expansion \kcdexpii\ do not correspond to the  two
parameters  in effective range theory;  indeed, we saw in the previous
section that the $\nu=0$ contribution already accounts for half of the
effective range.

By fitting the two free parameters $\tilde C_{\msb}$ and $\Lambda_\msb^2$ so
that the expression \kcdexpii\ correctly reproduces the \iso\ effective range
and scattering length, we arrive at the prediction for the phase shift plotted
as a solid line in fig.~7.
The values one finds for the parameters are now:
\eqn\relaii{
C_{\overline{MS}} (\mu)\bigg|_{\mu = m_{\pi}} = -{1\over (100\ \MeV)^2}\ ,
\qquad\qquad
{1\over \Lambda_{\overline{MS}}^2(\mu)} \bigg|_{\mu = m_{\pi}} =
-{1\over (121\ \MeV)^2}\ , }
which indicates a very significant improvement over those found by first
expanding the potential to order $\mu=1$ and then summing to all orders, eq.
\rela.  In particular, the momentum expansion scale $\Lambda$ is now much
larger.

Even with the larger scale $|\Lambda|=121\ {\rm MeV}$,
one would not expect the momentum expansion to converge
fast enough to be of use in nuclear matter, where
$p_F \sim 280\ {\rm MeV}$.
However, as argued in the previous section and evidenced by fig.~7, the
expansion for $\kcd$ has a much larger radius of convergence than the
derivative expansion in the Lagrangian.

\newsec{Conclusions}

We have shown how to perform a nonperturbative calculation of $NN$ scattering
in the \iso\ channel in an effective field theory expansion.  A key feature of
the procedure was the application of dimensional regularization
(usually viewed as a perturbative regulator)
and the $\msb$ renormalization scheme  procedure to the nonperturbative
problem.
Our results for the phase shift depend  only on physical observables and not
on any momentum cutoff, even though the bare $NN$ interactions in an EFT are
singular.

At leading order in the EFT expansion ($\nu=0$), which includes one pion
exchange and a contact interaction, we find a prediction for the effective
range \hbox{$r_0=1.3$ fm}, given the measured scattering length;  this is about
half the measured value.  The fit to the measured phase shift is poor above
\hbox{$|\bfp|\sim 25$ MeV}.  In order to better understand the range of
validity of the EFT approach, we investigated the phase shift including
effects at subleading order in the EFT expansion. At this order there is an
ambiguity about what quantity should be expanded; the ambiguity corresponds to
which higher order terms are kept in the EFT expansion to maintain unitarity.
 Following the method of \weinberg, one can expand the potential to subleading
order in the EFT expansion, and include its effects to all orders. Doing this,
we find the phase shift that results disagrees with data above
\hbox{$|\bfp|\sim 45$ MeV}, which is what one expects from the size of
coefficients one finds for the derivative expansion of the effective
Lagrangian.

An alternative method we explore is to expand the quantity $\kcd$ to subleading
order.  We explain why this expansion should be expected to have a greater
radius of convergence than the derivative expansion would lead one to expect,
at least at low orders in the EFT expansion.
This is supported by calculation, which suggests that the
\iso\ phase shifts at subleading order agree well with data at
up to \hbox{$\sim 280$ MeV}.
A strong correlation is implied between coefficients in the
derivative expansion of the Lagrangian that has to do with an $s$-channel pole,
but which remains to be quantitatively understood.

By investigating the EFT both with and without one pion exchange to order
$\nu=1$, we see that
including the pion increases the inverse mass scales that appear in the EFT
expansion, thereby improving improving its utility at high momentum.  Including
two pion exchange, four derivative interactions and possibly the effects of the
$\Delta$ will increase these
scales even further. Since higher partial waves are less sensitive to short
distance physics (and are in fact well approximated by one pion exchange, which
appears at lowest order in the EFT expansion), we are optimistic that the
techniques presented here will be successful at reproducing all of the spin
singlet partial wave phase shifts up to center of mass momenta comparable to
the Fermi momentum in nuclear matter.  This investigation is in progress.

As a consequence of our analysis we are forced to revise the power counting
scheme of Weinberg \weinberg\ in order to account for the powers of the nucleon
mass $M$ that appear in operator coefficients of the effective Lagrangian.  We
find that the EFT expansion is not equivalent to the chiral expansion, as the
EFT expansion requires summing contributions proportional to all powers of $M^2
m_\pi^2$.

Application of these techniques to the spin triplet channel is not
straightforward, however, since the interactions in this channel are singular
but not separable  (\eg, a $1/r^3$ singularity from one pion exchange).
Our hope is that this problem can be surmounted, in which case the techniques
we developed here should prove of use in a variety of interesting problems.
The EFT approach could be applied to nuclear matter, with the goal of
understanding its binding energy and compressibility  in terms of a few
parameters extracted from low energy scattering experiments. One could
investigate the implications of $SU(4)$ symmetry in $N$ and $\Delta$
interactions, recently shown to be a consequence of the large-$N_c$ expansion
of QCD
\ref\DaMan{
J.L. Gervais and B. Sakita, \prl{52}{1984}{87};
R. Dashen and A.V.~Manohar, \pl{315}{1993}{425}.}
\ref\KaSaa{D.B. Kaplan and M.J. Savage, \pl{365}{1995}{244}.}.
In particular, $SU(4)$ symmetry greatly reduces the number of four-fermion
operators one needs
to consider when the $\Delta$ is included \KaSaa .

Since $SU(3)$ flavor symmetry and its breaking can be easily incorporated in
the EFT formalism, it may prove a useful tool for exploring systems with
nonzero strangeness, extending the discussion of ref. \SaWia\  to a
nonperturbative analysis.
Finally, of great interest is the possibility   that the EFT analysis may prove
to be a useful tool in understanding systems at densities above nuclear
density, with an eye toward a systematic inclusion of nuclear forces in the
presently incomplete analyses of pion
condensation
\ref\BaCa{G. Baym and D.K. Campbell, in {\it Mesons and Nuclei}, edited by M.
Rho and D. Wilkinson, North Holland Pub. Co. (1979) 1031.}
and
kaon condensation~\ref\KaNea{D.B. Kaplan and A.E. Nelson,
\pl{175}{1986}{57}.}\nref\PoWi{
H. D. Politzer and M.B. Wise, \pl{273}{1991}{156};
D. Montano, H.D. Politzer and M.B. Wise,
\np{375}{1992}{507}.}--\ref\BrRh{G.E. Brown, C.-H. Lee, M. Rho and V. Thorsson,
Nucl. Phys. A567 (1994) 937;
C.-H. Lee, G.E. Brown, D.-P. Min and M. Rho, Nucl. Phys. A585 (1995) 401.}.

\vskip0.5in
\centerline{\bf  Acknowledgements}

We would like to thank G. Bertsch, A. Manohar, G. Miller, U. van Kolck, and R.
Venugopalan for useful
conversations and correspondence.
DBK would like to acknowledge a conversation with M.
Lutz on his related and independent work.
MJS was supported in part by the U.S. Department
of Energy under Grant No. DE-FG02-91-ER40682.
DBK was supported in part by DOE grant DOE-ER-40561, and NSF
Presidential Young Investigator award \pyidk.
MBW is supported in part by the Department of Energy
under Contract No. DE-FG03-92-ER40701.
\appendix{A}{Solving the Schr\"odinger equation}

In the \S3 we showed how to sum ladder diagrams involving a 4-nucleon contact
interaction.  This is formally equivalent to solving the Schr\"odinger equation
with a $\delta^3({\bf r})$ potential; however neither approach  makes sense
without
renormalization.  While the techniques of renormalization are familiar in the
context of field theory, here we show how to obtain the same results via the
Schr\"odinger equation
(See also ref. \ref\ericson{T. E. O. Ericson and L. Hambro, Ann. Phys. 107
(1977) 44.}.
This approach is quite convenient for practical
computations.

The equation  we want to solve is
\eqn\sequ{\eqalign{
0&=\[-{\nabla^2/ M} +  V_\pi({\bf r}) + \tilde{C}\delta^3({\bf r}) -
E\]\psi({\bf r})\cr
&\equiv \[H-E+\tilde{C}\delta^3({\bf r})\]\psi({\bf r})\ ,}}
where
\eqn\vpi{V_\pi({\bf r}) = -\alpha_\pi {e^{-m_\pi r}\over r}\ .}
Away from ${\bf r}=0$ we can find two independent $s$-wave solutions to
$(H-E)\psi=0$   We denote the regular $s$-wave solution by $\CJ_E(r)$ and the
irregular $s$-wave solution by $\CK^\lambda_E(r)$   They are normalized to have
the following
behaviour near $r=0$:
\eqn\orig{\eqalign{
\CJ_E(r)&\too{r\to 0} 1 - {\alpha_\pi M\over 2}r + \CO(r^2)\cr
\CK^\lambda_E(r)&\too{r\to 0} {M\over 4\pi r} -{\alpha_\pi M^2\over 4\pi}\ln
\lambda r + \CO(r\ln r)\ .}}
These functions have several features: \item{(i)} $\CK^\lambda_E$ is a Green's
function satisfying $$(H-E)\CK^\lambda_E=\delta^3({\bf r})\ ;$$
\item{(ii)} The arbitrary scale $\lambda$ in $\CK^\lambda_E$ corresponds to the
choice of boundary conditions on the Green's function (i.e,  the arbitrariness
in redefining $\CK_E(r)$ by an amount proportional to $\CJ_E(r)$);
\item{(iii)} For both functions, the dependence on the energy $E$ vanishes as
$r\to 0$; \item{(iv)} asymptotically, these functions become:
\eqn\asymprop{\eqalign{
\CJ_E(r)&\too{r\to \infty}  \(y {e^{ipr}/ pr} + c.c.\)\ ,\cr
\CK^\lambda_E(r)&\too{r\to \infty}  \(z {e^{ipr}/ pr} + c.c.\)\ ,}}
where $E=p^2/M$ and $y$ and $z$ are complex constants that must be determined
numerically.  However, $y$ and $z$ are related: since $\int \[\CK^\lambda_E
(H-E) \CJ_E\] {\rm d}^3r =0$, it follows upon integration by parts that
\eqn\yzrel{ y z^* - y^*z = -{ipM\over 8\pi}\ .}

The Schr\"odinger equation \sequ\ can now be rewritten as
\eqn\schroAone{
(H-E)\psi({\bf r}) = -\tilde{C}\psi({\bf 0})\delta^3({\bf r})
 \ ,}
 and is formally solved in the s-wave channel by
\eqn\solui{
\psi({\bf r}) = a \CJ_E(r) + b\CK^\lambda_E(r)\too{r\to\infty}
(ay+bz){e^{ipr}\over pr} + c.c.}
provided that
$b=-\tilde C\psi({\bf 0})$,
or
\eqn\soluii{
b=-a{\tilde{C}\over 1+ \tilde{C} \CK^\lambda_E(0)}\ .}
(We say ``formally'' since $\CK^\lambda_E(0)$ is divergent;  we will address
the issue of renormalization below).
Note that the ratio $a/b$ is real.
Comparing eq. \solui\ with the desired ($s$-wave) asymptotic boundary condition
\eqn\desired{
\psi({\bf r}) \too{r\to\infty} -\frac{i}{2}\( e^{2i\delta} {e^{ipr}\over pr} -
{e^{-ipr}\over pr}\)
\ ,}
it follows that the phase shift is given by
\eqn\eid{\openup 2 \jot \eqalign{
e^{2i\delta} &= -{(ay+bz)/ (ay+bz)^*}\cr
&= -{y\over y^*} - \({1\over y^*}\)^2{zy^*-yz^*\over a/b + z^*/y^*}
\cr
&= e^{2i\delta_\pi} -  \({1\over y^*}\)^2\({ipM\over 8\pi}\) {1\over
-1/\tilde{C} -
\CK^\lambda_E(0)+z^*/y^*}\ \ ,}}
where we have made use of eqs. \yzrel, \ \soluii, and have defined $\delta_\pi$
to be the ``OPE'' s-wave phase shift arising from the one pion exchange Yukawa
interaction $V_\pi$, and no contact term ($\exp(2i\delta_\pi)=-y/y^*$) (see
fig.~8) \foot{Our numerical calculations were performed with $m_\pi=140\ \MeV$
and $M=940\ \MeV$.}.
\topinsert
\centerline{\epsfxsize=3in\epsfbox{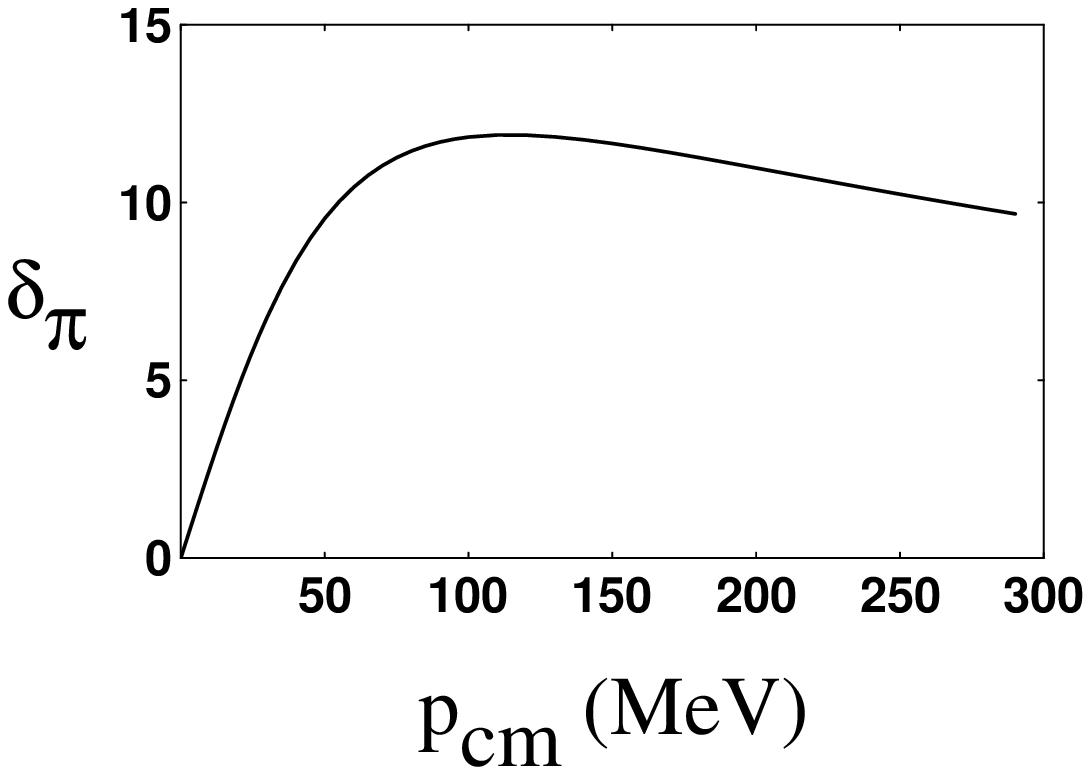}}
\smallskip
\caption{Fig. 8.  The OPE phase shift $\delta_\pi$ (in degrees) as a function
of $|\bfp|$ in MeV.
}
\endinsert

It is
now just a few steps to relate the above expression to eq. \famp, the analogous
formula derived diagrammatically.  First note that the canonically normalized
scattering solution in the pure Yukawa theory is given by  $\chi_{\bf p}({\bf
r}) =
-i\CJ_E(r)/(2y^*)$, so that
\eqn\chizeroap{
\chi_{\bf p}({\bf 0})= -i/(2y^*)
\ .}
The function $\left[ \chi_{\bf p}({\bf 0})\right]^2$ is plotted in fig.~9 as a
function
of the centre-of-mass momentum.
\topinsert
\centerline{\epsfxsize=3in\epsfbox{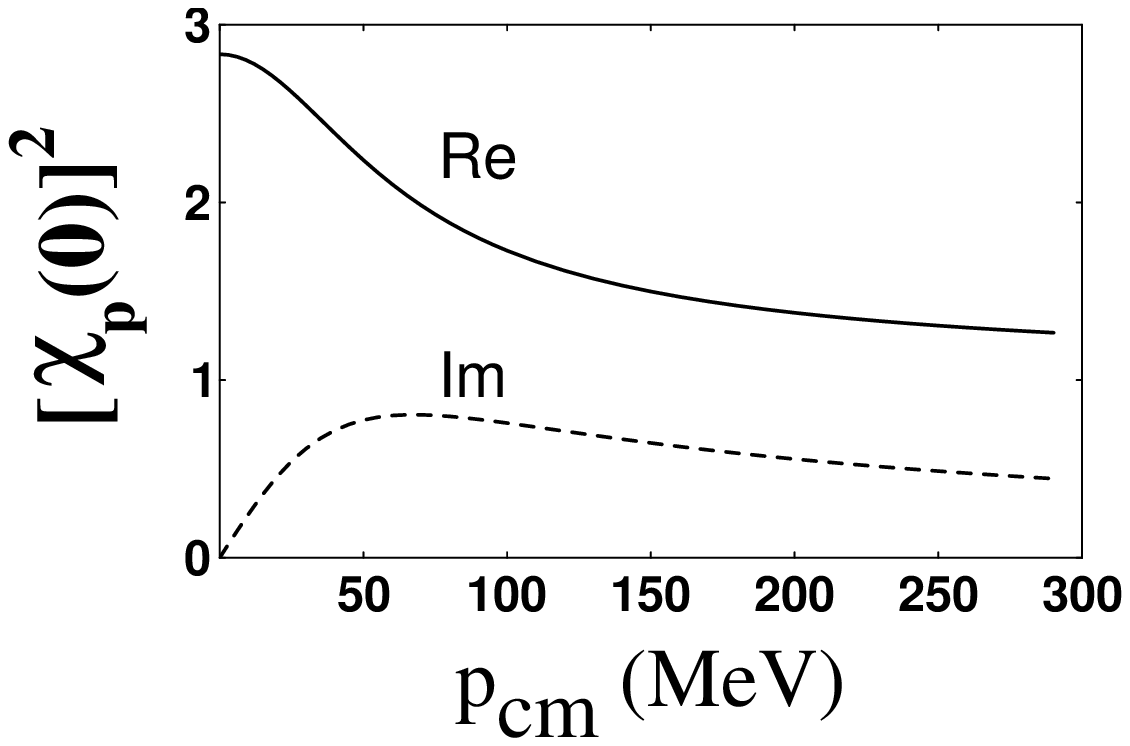}}
\smallskip
\caption{Fig. 9.  The OPE wave function squared at the origin,
$[\chi_{\bfp}({\bf 0})]^2$, plotted versus momentum in MeV.  The solid and
dashed lines correspond to the real and imaginary parts respectively.}
\endinsert
Next note that the
retarded Green's function (satisfying the asymptotic boundary condition that
there is no incoming wave) is given by
\eqn\gtilrzero{
\tilde{G}_E({\bf r},{\bf 0}) = (-\CK_E^\lambda(r) +
(z^*/y^*)\CJ_E(r))
 \  ;}
finally, the relation between the Feynman amplitude
$i\CA$ and the phase shift is
\eqn\iAmpfeyn{
i\CA = i{4\pi(e^{2i\delta}-1)\over 2ipM}
 \  .}
It follows that expression \eid\ is equivalent to eq. \famp:
\eqn\iCfeyn{
i\CA = i\CA_\pi -i{\tilde C \[\chi_{\bf p}({\bf 0})\]^2\over 1 - \tilde C
\tilde{G}_E ({\bf 0,0})}
 \ .}
When the effects of pions are not included, as in \S3, one recovers the
amplitude \ert, since $\alpha_\pi\to 0$ implies $\CA_\pi\to 0$, $\tilde C\to
C$, $\chi_{\bfp}({\bf 0})\to 1$, and $G_E\to G_E^0$ in the above expression.

So far the discussion has been in terms of the  quantity, $\CK^\lambda_E(0)$,
which was seen in eq. \orig\ to have both linear and logarithmic divergences as
$r\to 0$.  This can be remedied by renormalizing $\tilde{C}$, for example by
defining
\eqn\rcval{{1\over \tilde{C}_R(\lambda)}\equiv {1\over \tilde{C}} +
\CK^\lambda_0(0)={1\over
\tilde{C}} + \CK^\lambda_E(0)\ ,}
(where we used the fact that $\lim_{r\to 0} [\CK^\lambda_E(r) -
\CK^\lambda_0(r)] = 0$, from eq. \orig).
Then eq. \eid\  for the phase shift can be written in terms of renormalized
quantities as
\eqn\reid{e^{2i\delta}= e^{2i\delta_\pi} -  \({1\over y^*}\)^2\({ipM\over
8\pi}\) {1\over -1/\tilde{C}_R(\lambda) +z^*/y^*}\ .}
This choice of renormalization scheme is convenient for computations:
First
one solves for the solutions to the Schr\"odinger equation with the  OPE
potential $V_\pi({\bf r})$,  subject to the boundary condition \orig;  from
that one
computes the asymptotic behaviour and the coefficients $y$ and $z$ in eq.
\asymprop;  then one computes $C_R(\lambda)$ from eq. \reid\ in terms of the
measured \iso\ scattering length $a=-23.7$ fm.  Note that this expression is
very
different than the pure OPE calculation.

The renormalization prescription \rcval\ is different than \ms,  but it is
straightforward to relate the two.  Using
\eqn\rela{ {1\over\tilde{C}} - \tilde{G}_E ({\bf 0,0}) =
{1\over \tilde{C}_{\overline{MS}}} -
\tilde{G}_E^{\overline{MS}}({\bf 0,0})\ , }
one finds
\eqn\relb{\eqalign{ {1\over \tilde{C}_{\overline{MS}}} &=
{1\over \tilde{C}} - [\tilde{G}_E({\bf 0,0}) -
\tilde{G}_E^{\overline{MS}}({\bf 0,0})] \cr
&= {1\over \tilde{C}_R(\lambda)} -\CK_0^\lambda (0) - [\tilde{G}_0 ({\bf 0,0})
-
G_0^{\overline{MS}} ({\bf 0,0})]\ .}}
Here we used the fact that the difference $[\tilde{G}_E ({\bf 0,0}) -
\tilde{G}_E^{\bar{MS}} ({\bf 0,0})]$ is independent of $E$.
Rearranging eq. \relb\
using the fact that only the first two diagrams in the perturbative expansion
of $\tilde{G}_0 ({\bf 0,0})$ are divergent yields
\eqn\relc{\eqalign{
{1\over \tilde{C}_{\overline{MS}}} &=
{1\over \tilde C_R (\lambda)}
- \lim_{r' \rightarrow 0}
\left[
\CK_0^\lambda (r')
+ \langle {\bf r}' | \hat G_0^0|{\bf r} = {\bf 0}\rangle
+ \langle {\bf r}'| \hat G_0^0 \hat V_\pi \hat G_0^0| {\bf r} = {\bf 0}\rangle
\right] \cr
& + \left[
\langle {\bf r}' = {\bf 0} | \hat G_0^0 | {\bf r} = {\bf
0}\rangle_{\overline{MS}}
+ \langle {\bf r}' = {\bf 0}| \hat G_0^0 \hat V_\pi \hat G_0^0| {\bf r} = {\bf
0}
\rangle_{\overline{MS}}\right]\ . }}
Explicit calculation gives
\eqn\reld{\eqalign{
\langle {\bf r}'|\hat G_0^0 | {\bf r} = {\bf 0}\rangle &= {-M\over 4\pi r'}\ ,
\cr
\langle {\bf r}'| \hat G_0^0 \hat V_\pi \hat G_0^0 | {\bf r} = {\bf 0}\rangle
&= - {\alpha_\pi
M^2\over 4\pi} [1 - \ln(m_\pi r') - \gamma +\ldots]\ , }}
where the ellipses represent terms that vanish as $r' \rightarrow 0$.  In eq.
\reld\ $\gamma$ is Euler's constant $(\gamma \simeq 0.577)$.
Using dimensional regularization
\eqn\rele{ \langle {\bf r}' = {\bf 0} | \hat G_0^0| {\bf r} = {\bf 0}
\rangle_{{DIM\atop REG}} = 0}
and
\eqn\relf{ \langle {\bf r}' = {\bf 0} | \hat G_0^0 \hat V_\pi \hat G_0^0
| {\bf r} = {\bf 0}
\rangle_{{DIM\atop REG}} = - 4\pi \alpha_\pi M^2 I_n (m_\pi)\ , }
where $I_n(m_\pi)$ is the two loop integral
\eqn\relg{ I_n(m_\pi) = \int {d^n {\bf q}\over (2\pi)^n}
\int {d^n {\bf k}\over (2\pi)^n}
{1\over {\bf q}^2} {1\over {\bf k}^2} {1\over [({\bf q}-{\bf k})^2 + m_\pi^2]}\
.}
Combining denominators with the Feynman trick
\eqn\relh{\openup2\jot\eqalign{
I_n (m_\pi) &= \int_0^1 dx \int {d^n {\bf q}\over (2\pi)^n {\bf q}^2}
\int {d^n {\bf k}\over (2\pi)^n}
{1\over [{\bf k}^2 + {\bf q}^2 x (1 - x) + m_\pi^2 x]^2} \cr
&= \int_0^1 dx \int {d^n {\bf q}\over (2\pi)^n {\bf q}^2}
{\pi^{n/2}\over (2\pi)^n}
{\Gamma (2 - n/2)\over [{\bf q}^2 x (1 - x) + m_\pi^2 x]^{2 - n/2}}\ . }}
Changing the momentum integration variable from ${\bf q}$ to
${\bf p} = \sqrt{1-x}{\bf q}$ the above becomes
\eqn\reli{ I_n (m_\pi) = {\pi^{n/2} \Gamma(2 - n/2)\over (2\pi)^{2n}}
\int_0^1 dx x^{n/2 - 2} (1-x)^{1 - n/2}
\int {d^n {\bf p}\over {\bf p}^2} {1\over [{\bf p}^2 + m_\pi^2 ]^{2 - n/2}}\ .}
Performing the ${\bf p}$ integration and the $x$ integration gives
\eqn\relj{ I_n (m_\pi) = - {\pi^n \Gamma(3 - n) (m_\pi^2)^{n - 3}\over
(2\pi)^{2n}} \Gamma \left({n\over 2} - 1\right)
\Gamma \left(1 - {n\over 2}\right)\ . }
Consequently in the $\overline{MS}$ subtraction scheme
\eqn\relk{\eqalign{
\langle {\bf r}' = {\bf 0}| \hat G_0^0 | {\bf r}
= {\bf 0}\rangle_{\overline{MS}} &= 0, \cr
\langle {\bf r}' = {\bf 0}| \hat G_0^0 \hat V_\pi \hat G_0^0|
{\bf r}  = {\bf 0}\rangle_{\overline{MS}}
&= -{\alpha_\pi M^2\over 8\pi} \left[1 - \ln \left({m_\pi^2\over
\mu^2}\right)\right]. }}
Combining these results and using the small $r$ behavior of
$\CK_0^\lambda (r)$ in eq. \orig\ yields
\eqn\rell{ {1\over\tilde{C}_{\overline{MS}}} =
{1\over \tilde{C}_R(\lambda)} - {\alpha_\pi
M^2\over 8\pi} \left[\ln \left({\mu^2\over\lambda^2}\right)
+ 2\gamma - 1\right]
 \  ,}
and consequently,
\eqn\Gemsbardef{
G_E^{\msb} ({\bf 0,0}) =
{z^*\over y^*}
\  -\   {\alpha_\pi
M^2\over 8\pi} \left[\ln \left({\mu^2\over\lambda^2}\right) + 2\gamma -
1\right]
\ .}
The $\lambda$ dependence of $1/\tilde C_R (\lambda)$ is exactly cancelled by
the $\lambda$ dependence of $\ln \left({\mu^2\over\lambda^2}\right)$
appearing in the second term on the r.h.s of \rell\ , leaving
$ 1/\tilde{C}_{\overline{MS}}$ independent of $\lambda$.
Similarly, cancellation of the $\lambda$ dependent terms on the r.h.s. of
\Gemsbardef\ leaves $G_E^{\msb}({\bf 0,0})$ independent  of  $\lambda$.
$G_E^{\msb}({\bf 0,0})$ is plotted in fig.~10 for a subtraction point of
$\mu = m_\pi$.

\topinsert
\centerline{\epsfxsize=3in\epsfbox{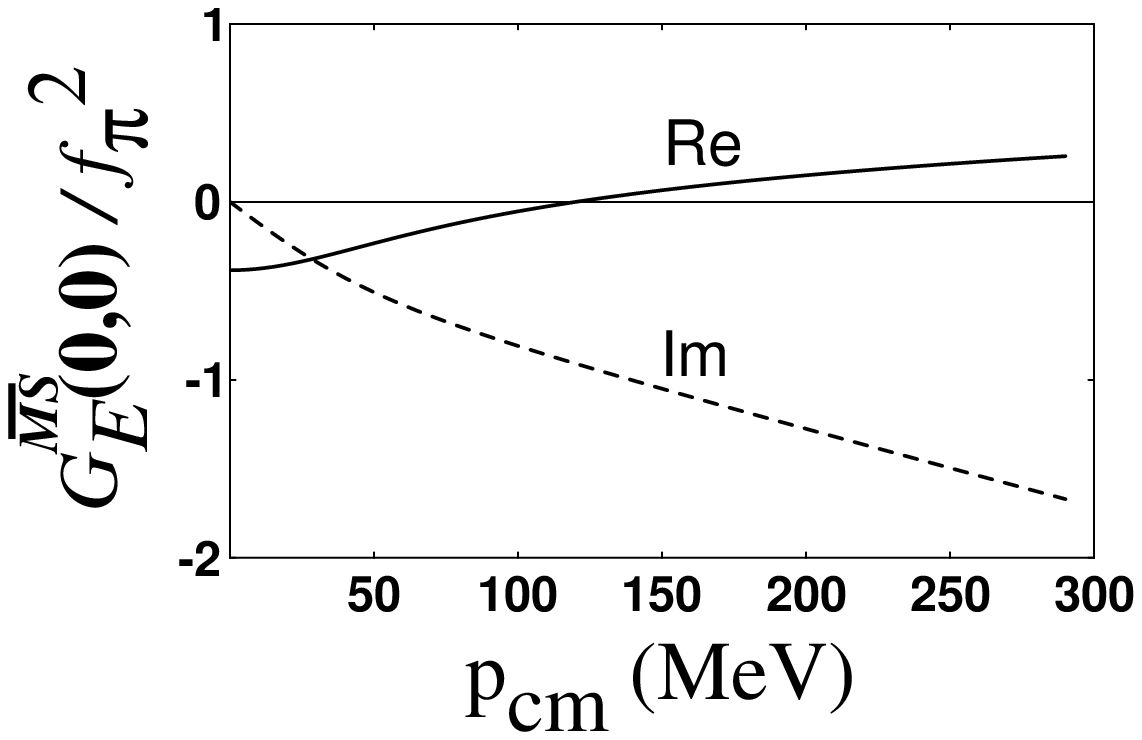}}
\smallskip
\caption{Fig. 10.  The $\msb$ Green function at the origin normalized to
$f_\pi^2$, $G_E^{\msb}({\bf 0,0})/f_\pi^2$, as a function of $|\bfp|$ in MeV.
The solid and dashed lines are the real and imaginary parts, respectively.}
\endinsert

\appendix{B}{Computing the effects of 2-derivative, 4-nucleon interactions}

In this appendix we show how to compute the sum of ladder diagrams in \ms\ when
a two-derivative, 4-nucleon operator is included.  Consider the ladder sum
including both the contact interactions \vertval\ as well as the Yukawa part of
one pion exchange:
\eqn\potsum{
\hat V= \hat V_\pi + \hat V_c
\ ,}
where $\hat V_\pi$ is given in eq. \cydef\ as
\eqn\potaldef{
\bra{\bfp}\hat V_\pi\ket{\bfp'} = -{4\pi\alpha_\pi\over
(\bfp-\bfp')^2+m_\pi^2} \ ,\qquad \alpha_\pi = {g_A^2 m_\pi^2\over 8 \pi
f_\pi^2}\ ,}
and $f_\pi=132\ \MeV$ is the pion decay constant. (For the effective theory
without pions, one can take the result we will derive and set $\alpha_\pi=0$).
The contact interaction $\hat V_c$ is given by,
\eqn\potmattwo{
\bra{\bfp}\hat V_c\ket{\bfp'} = \tilde C\(1+{\bfp^2 + \bfp'^2\over 2
\Lambda^2}\)
 \  .}
This can be conveniently rewritten as
\eqn\vertvalii{V_c(\bfp,\bfpp)=  \tilde{C} \(1+  ME/\Lambda^2\) - \tilde{C}
M\({(E-\bfp^2/M) + (E-\bfpp^2/M)\over 2 \Lambda^2}\)\ , }
or in operator form as
\eqn\vertvaliii{\hat V_c = \tilde{C}
\dint {{\rm d}^3{\bfq}\over (2\pi)^3} {{\rm d}^3{\bfq'}\over (2\pi)^3}
\((1+ME/\Lambda^2)\,\ket{\bfq}\bra{\bfq'}- \biggl\{
M(\hat G_E^0)^{-1}/\Lambda^2,\ket{\bfq}\bra{\bfq'}\biggr\}\)}
where $(\hat G_E^0)^{-1}=(E-\hat H_0)$.

The first term in eq. \vertvalii\  is easy to deal with, since $E$ is a number,
not an operator.  Computing the effects of the second term  involving the
operator $ (\hat G_E^0)^{-1}$ requires some thought.
There are three ways the $ (\hat G_E^0)^{-1}$ term can enter the ladder diagram
calculation.

Firstly, there could be an insertion acting on  the ``external'' legs (by
external legs we mean nucleon propagators which interact via $\hat V_\pi$, but
not
through any contact interactions).  This entails calculating the integral
($\bra{\bfp}$ is an on-shell state)
\eqn\drvi{\eqalign{\igraln{\bfq}\bra{\bfp} (\hat G_E^0)^{-1}\hat G_E
(\hat G_E^0)^{-1}\ket{\bfq}&=
\igraln{\bfq}\bra{\bfp} (\hat G_E^0)^{-1}(1+\hat G_E \hat V_\pi)\ket{\bfq}\cr
&= \alpha_\pi m_\pi \igraln{\bfq}\bra{\bfp} (\hat G_E^0)^{-1}\hat G_E
\ket{\bfq}\ ,}}
where we made use of the fact that $\bra{\bfp} (\hat G_E^0)^{-1}\ket{\bfq}=0$
for an
on-shell state $\bra{\bfp}$; as well as the dimensionally regulated integral
\eqn\drvii{\igraln{\bfq}\bra{\bfq'}\hat V_\pi\ket{\bfq} =
\igraln{\bfq}{-4\pi\alpha_\pi\over (\bfq-\bfq')^2+m_\pi^2} = \alpha_\pi
m_\pi\ .}
This is equivalent making the replacement $(\hat G_E^0)^{-1}\to \alpha_\pi
m_\pi$.
(Only the second factor of $(\hat G_E^0)^{-1}$ in eq. \drvi\  comes from the
interaction $\hat V_c$; the first  $(\hat G_E^0)^{-1}$ is there to amputate the
outgoing propagator).

Secondly,  one insertion of $(\hat G_E^0)^{-1}$ could act on internal nucleon
lines
(dressed by $\hat V_\pi$).  This gives rise to the integral
\eqn\drviii{\eqalign{
\digraln\bra{\bfq} \hat G_E (\hat G_E^0)^{-1}\ket{\bfq'}
&=
\digraln\bra{\bfq}(1+\hat G_E\hat V_\pi)\ket{\bfq'}\cr
& =
\alpha_\pi m_\pi\digraln\bra{\bfq} \hat G_E \ket{\bfq'}\ ,}}
where we made use of the relation \lad\ between the full and free propagators,
of the integral \drvii, and of the fact that $\igraln{q} \bfq^{2r}=0$ in
dimensional regularization.   Again, $(\hat G_E^0)^{-1}$ just gets replaced by
$\alpha_\pi m_\pi$.

Finally, two insertions of $(\hat G_E^0)^{-1}$ could act on internal lines:
\eqn\drviii{\eqalign{
\digraln&\bra{\bfq} (\hat G_E^0)^{-1}\hat G_E (\hat G_E^0)^{-1}\ket{\bfq'}
 \cr &=
\digraln\bra{\bfq}(\hat G_E^0)^{-1}(1+\hat G_E  \hat V_\pi)\ket{\bfq'}\cr
& =
\alpha_\pi m_\pi\digraln\bra{\bfq} (\hat G_E^0)^{-1}\hat G_E
\ket{\bfq'}\cr
&=
(\alpha_\pi m_\pi)^2\digraln\bra{\bfq} \hat G_E \ket{\bfq'}\ .}}
Again we see that $(\hat G_E^0)^{-1}$  gets replaced by $\alpha_\pi m_\pi$.
In conclusion, given eqs. \vertvalii, \vertvaliii,  the effect of including the
2-derivative operator in eq. \lagi\ is simply to replace $\tilde C$ by
\eqn\corep{\tilde C\to \tilde C\(1-{\alpha_\pi m_\pi M\over \Lambda^2}
\ +\  {E M\over \Lambda^2}\)\ .}
This is the result utilized in \S4.  The expression \isoampii\ in \S3
involves the substitution \corep\ with $\alpha_\pi$ set to zero (no pion
contribution).

\listrefs

\bye